\documentclass[journal=jacsat,manuscript=article]{achemso}
\pdfoutput=1
\usepackage{setspace}
\usepackage{graphicx}
\usepackage{rotating}
\usepackage{amsmath, amssymb, amsthm}
\usepackage{color}
\usepackage{hyperref}
\author{Jeseong Yoon}
\affiliation[Korea Institute for Advanced Study]
{School of Computational Sciences, Korea Institute for Advanced Study, Seoul 130-722, Korea}
\author{D. Thirumalai}
\affiliation[University of Maryland, Korea Institute for Advanced Study]
{Institute for Physical Sciences and Technology, University of Maryland, College Park 20742, USA}
\author{Changbong Hyeon}
\email{hyeoncb@kias.re.kr}
\affiliation[Korea Institute for Advanced Study]
{School of Computational Sciences, Korea Institute for Advanced Study, Seoul 130-722, Korea}

\title[Urea-induced denaturation of nucleic acids]
{Urea-induced denaturation of preQ$_1$-riboswitch}

\begin{document}
\begin{abstract}
Urea, a polar molecule with a large dipole moment, not only destabilizes the folded RNA structures, but can also enhance the folding rates of large ribozymes. Unlike the mechanism of urea-induced unfolding of proteins, which is well understood, the action of urea on RNA has barely been explored. 
We performed extensive all atom molecular dynamics (MD) simulations to determine the molecular underpinnings of urea-induced RNA denaturation.  
Urea displays its denaturing power in both secondary and tertiary motifs of the riboswitch (RS) structure.   
Our simulations reveal that the denaturation of RNA structures is mainly driven by the hydrogen bonds and stacking interactions of urea with the bases. 
Through detailed studies of the simulation trajectories, we found that geminate pairs between urea and bases due to hydrogen bonds and stacks persist only $\approx (0.1-1)$ ns, which suggests that urea-base interaction is highly dynamic.   Most importantly, the early stage of base pair disruption is triggered by penetration of water molecules into the hydrophobic domain between the RNA bases.  
The infiltration of water into the narrow space between base pairs is critical in increasing the accessibility of urea to transiently disrupted bases, thus allowing urea to displace inter base hydrogen bonds.    
This mechanism, water-induced disruption of base-pairs resulting in the formation of a "wet" destabilized RNA followed by solvation by urea, is the exact opposite of the two-stage denaturation of proteins by urea. In the latter case, initial urea penetration creates a dry-globule, which is subsequently solvated by water penetration leading to global protein unfolding. Our work shows that the ability to interact with both water and polar, non-polar components of nucleotides makes urea a powerful chemical denaturant for nucleic acids.  
\end{abstract}

\section{Introduction}
Neutral in charge, but with a large dipole moment (4.56 Debye) and a size comparable to water,  
urea is a versatile chemical agent that can interact with both polar and non-polar components of biopolymers \cite{Roseman75JACS}. 
Because of its ability to denature folded states, urea has long been used to study the thermodynamic stability of proteins and more recently RNA, and infer the kinetic mechanisms of their folding from urea-dependent folding and unfolding rates \cite{KiefhaberJMB03_1,shelton1999biochemistry}.
Recent experimental studies using hydrogen exchange \cite{Lim09PNAS}, mid-infrared pump-probe spectroscopy probing the orientational mobility of water molecule \cite{rezus2006PNAS} have provided additional insights into the molecular mechanisms of how urea destabilizes proteins. 
In addition, molecular simulations have played a critical role in elucidating the microscopic mechanism of urea-induced denaturation \cite{Kuharski84JACS,Caflisch99Structure,Mountain03JACS,obrien2007JACS, Stumpe07JACS,hua2008PNAS,Canchi10JACS}. 
Given the complexity of water dynamics that is critical in determining the effective forces between solutes in condensed media \cite{Chandler05Nature}, it is remarkable that the hydrogen bond network of aqueous water is relatively unperturbed by urea even at high concentrations (Figure S1). 
This unique property of urea rules out the ``indirect interaction model" as a plausible mechanism for the urea-induced denaturation \cite{rezus2006PNAS,obrien2007JACS,hua2008PNAS}.  
Although the details of the energetics associated with urea-protein interactions are still under active debate, MD simulations have unambiguously shown that urea molecules act thermodynamically like a ``surfactant"; first, water molecules in the first solvation shell (FSS) are expelled by making polar interaction (hydrogen bond) with the surface exposed amide backbone; later intrude into the protein core by interacting with nonpolar hydrophobic and polar side chains \cite{hua2008PNAS}.
 
More recently, there is considerable interest in understanding how urea and naturally occurring osmolytes (TMAO, betaine, etc) affect the conformations of RNA \cite{Lambert07JMB,Pincus08JACS}. It is natural to expect that just like proteins, whose folded and unfolded population can be manipulated by cosolvents as well as temperature, mechanical force, and pressure, RNAs also adapt to changing cosolvent environment \cite{Lambert07JMB}. 
Indeed, with the demonstration that modest amount of urea can promote folding of misfolded, kinetically trapped structures of \emph{Tetrahymena} ribozyme urea has also been used as reagent to probe RNA folding \cite{Pan97JMB,PanPNAS99,WilliamsonJMB98,SosnickNSB97}.  
However, unlike the relatively well understood action of urea on proteins, detailed molecular mechanism of urea-induced denaturation of nucleic acids has not been explored.  
Because of the differences in the basic building block between RNA and proteins \cite{Thirum05Biochem} it is unclear if the urea-induced denaturation mechanism of proteins and RNA should be similar.  
The physicochemical property of the phosphodiester backbone of RNA differs fundamentally from that of polypeptide chain of proteins. 
%Unlike flexible protein backbone that use hydrogen bondings to form $\alpha$-helices or $\beta$-sheets, RNA backbone is not free to form intra-backbone hydrogen bonds. 
Although seven torsion angles along the phosphodiester backbone leads to a diversity of the RNA configurations \cite{Murray03PNAS,Wadley07JMB}, RNA backbone itself does not actively participate in stabilizing the RNA structure. 
The stability of RNA mainly arises from inter base (base pairing as well as stacking) and tertiary interactions.  
These considerations and the ease with which RNA can adopt alternative structures raise the possibility that interaction of urea on RNA with tertiary fold can be far more complex than that found in proteins. 

Previously we established, using all-atom MD simulation, that urea disrupts RNA base pair interactions by forming multiple hydrogen bonds and stacking interaction, thus potentially serving as a surrogate for a base \cite{Priyakumar09JACS}. 
However, our work was restricted to urea effects on RNA secondary structures.  Here, we study the dynamic aspects of how urea disrupts tertiary and secondary interactions in RNA using a  36-nt \textit{Bacillus subtilis} preQ$_{1}$-riboswitch aptamer domain as a model system. Despite its relatively small size, preQ$_{1}$-RS, containing both secondary and tertiary motifs arranged in a characteristic H-type pseudoknot structure \cite{Zhang11JACS}, is an excellent system for probing the mechanism of urea-induced RNA denaturation from multiple angles.  
We generated multiple $600-700$ ns 
%(in total 9.2 $\mu sec$) 
MD simulation trajectories of preQ$_{1}$-RS at three urea concentrations (0 M, 4 M, 8 M) (see Figures \ref{Fig1}b, \ref{Fig1}c and Figure S2). 
%Although the RS was not fully denatured even in 8 M urea on the time scale of each trajectory ($\lesssim$ 1 $\mu$s), we could observe partial disruptions and rearrangements of the RNA structure in several places, which is sufficient to propose a mechanism for urea action on RNA.  
%Given that the biological function of preQ$_{1}$-RS, i.e., the translational gene regulation, is dictated by the metabolite-dependent conformational changes accompanied with secondary structure rearrangement \cite{AlHashimi08COSB,rieder2010PNAS}, how RS reassembles in response to the chemical stress would be of great interest. Instead, removal of the preQ$_1$ and Ca$^{2+}$, which strongly stabilize the RS structure \cite{Zhang11JACS} by binding at the core region of RS (P1 helix and L1, L2 loops. See Fig.\ref{Fig1}a), induces a significant rearrangement in both secondary and tertiary structures (see SI text and Fig.S3a).
 Analyses of the trajectories from our all-atom MD simulations have allowed us to discern the microscopic mechanism of urea-induced denaturation and accompanying role of water dynamics. 
First, as in the earlier study \cite{Priyakumar09JACS}, but here based on much longer simulation time scale, we show that the major changes in urea and water density occur around nucleobases, which is revealed by visualization of urea and water at early and late stages of the simulation.  
Second, we study how urea interactions with bases destabilize the secondary and tertiary motifs in the RS structure. 
Third, we study the kinetics and energetics of urea-base stacking and hydrogen bond interactions in detail. 
Our analyses show that compared to the time scale of global denaturation process, the lifetimes of urea forming base stacks and hydrogen bonds are relatively short $\lesssim\mathcal{O}(1)$ ns, which indicates that the denaturation of RNA molecule is the consequence of an accumulation of numerous attempts to disrupt intramolecular base-base interactions.  
These highly dynamic interactions between urea, water, and nucleic acids are exploited to disrupt the inter base hydrogen bonds and stacks and to stabilize the isolated nucleobases. From a kinetic perspective, we find that in the early stages there is an increase of water population around the bases before the urea molecules start interacting with nucleobase.   
In other words, the bases are initially destabilized by penetration of water molecules, creating a destabilized "wet" RNA. Only subsequently does urea solvate the solvent exposed bases. 
This finding underscores the critical role of water at the early stage of base pair disruption. 
The findings from this study, which shed light on the microscopic dynamics of urea and water in the process of RNA denaturation, are amenable to experimental scrutiny.  

\section{Results and Discussions}
\begin{figure}[h!]
\centering
\includegraphics[scale=0.21]{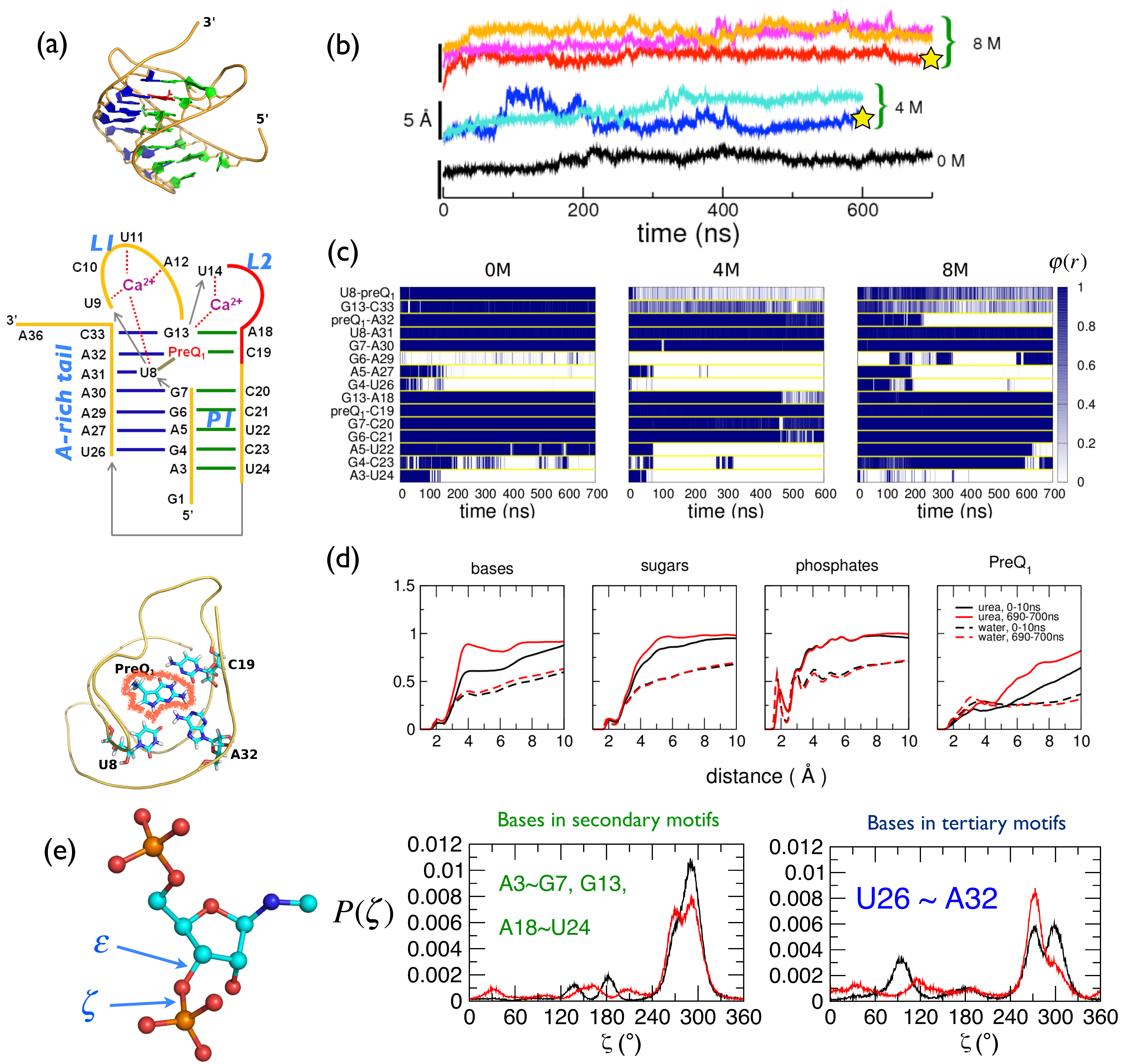} 
\caption{Effect of urea on riboswitch structure. {\bf (a)} Structure of preQ$_{1}$-riboswitch. 
Diagram below shows base pairs in the RS, with green lines for secondary and blue lines for tertiary interactions. 
The metastable preQ$_{1}$ lies at the center of RS structure, mediating both secondary and tertiary interaction. 
\textbf{(b)}
Time evolution of RMSD of RS.  
\textbf{(c)} Time evolutions of base pair contacts at 0 M, 4 M, and 8 M urea concentration. 
The evolutions of contacts for 4 and 8 M were calculated using the trajectories marked with star in (b). 
We used $\varphi(r)$ defined in Figure \ref{eqn:logistic} in the main text to visualize the state of each base pair contact. 
The scale for $\varphi(r)$ is on the right. 
\textbf{(d)} Radial distribution functions of water (dashed line) and urea (solid line) around RNA base, sugar, and phosphate of RS and preQ$_1$ calculated at early ($0-10$ ns) and late stage ($690-700$ ns) using 8 M urea trajectory (the rightmost panel in Figure 1c).
Here, the distance $r$ is the minimum distance from urea (or water) to any heavy atom in the specified group (base, sugar, phosphate, preQ$_1$). 
{\bf (e)} Distributions of angle $\zeta$ for the bases in secondary and tertiary motifs calculated for the first (black) and last (red) 10 ns simulation in 8 M urea. 
}
\label{Fig1}
\end{figure}

\noindent {\bf Effect of urea on the RS structure.}
A cognate metabolite preQ$_1$ bound to the ligand binding site of the preQ$_1$ riboswitch consolidates the RS into a characteristic H-type pseudoknot structure, the core part of which is stabilized by the tertiary zipper contacts between P1 helix (G1$-$G7, C20$-$U24) and A-rich tail (A25$-$A32) (see Figure \ref{Fig1}a). 
PreQ$_1$ ligand bound to the center of the P1 helix acts as a surrogate nucleotide, stabilizing both the secondary and tertiary structures by forming hydrogen bonds or stacks with five neighboring nucleotides: three in-plane interactions with U8, C19, A32, and two stacks with G7 and G13.

In the absence of urea, most of the base pairs identified in the NMR structure \cite{Zhang11JACS} remain intact throughout the 700 ns simulation except the Watson-Crick (WC) G4-C23 base pair located at the end of P1 helix, the disruption of which leads to an increase of RMSD by $\sim$2 \AA\ (0 M trajectory in Figure \ref{Fig1}b). 
Addition of urea to solution destabilizes the RS structure. 
The overall destabilization of the RS structure at 0 M, 4 M, and 8 M urea is visualized using 
time evolutions of RMSD (Figure \ref{Fig1}b) and base-pair contacts involving both the secondary and tertiary structures (Figure \ref{Fig1}c, Figure S2).   
To make clear distinction between the bound and  unbound states of base pairs in Figure \ref{Fig1}c and Figure S2, we used a logistic function,  
\begin{equation}
\varphi(r)=\frac{1}{1+e^{(r-r_0)/\sigma}}
\label{eqn:logistic}
\end{equation}
where $r$ is a distance between nitrogen (N) or oxygen (O) atom and its partner heavy atom participating in base pair hydrogen bond. 
We chose $r_0=5.5$ \AA\ and $\sigma=0.33$ \AA\ by considering the mean and standard deviation of base pair distances in the intact state.
Interestingly, many of preQ$_{1}$-nucleobase pairs (preQ$_1$-A32, preQ$_1$-C19) remain intact in our simulation trajectories lasting $\approx 700$ ns even at 8 M urea (Figure \ref{Fig1}c, Figure S2). 
As a result, we failed to observe the global denaturation of RS structure via disruption of interactions associated with preQ$_1$. 
Although 700 ns is clearly not long enough to induce full denaturation of RS (see, however, the Supplementary Information (SI) text and Figure S3 for the dynamics of RS upon removal of preQ$_1$ and Ca$^{2+}$ showing larger scaling unfolding in the absence of the ligand), the dynamics of RS structure in aqueous urea displays many interesting features, which shed light on the action of urea on RNA.      
\\

\noindent \textit{Urea and water radial distribution functions (RDFs) around RS.}
The change in solution environment around nucleic acid structure reflects the denaturation of RS. 
RDFs of urea ($\mathrm{g_u}(r)$) and water ($\mathrm{g_w}(r)$) around base (B), sugar (S), phosphate (P), and preQ$_1$ (Q), which we denote as $\mathrm{g}_x(r|\mathrm{B})$, $\mathrm{g}_x(r|\mathrm{S})$, $\mathrm{g}_x(r|\mathrm{P})$, and $\mathrm{g}_x(r|\mathrm{Q})$, respectively, with $x=$w or u, and where the distance $r$ is the minimum distance from urea (or water) to any heavy atom in the specified group, provide quick glimpse into the solvent structure (Figure \ref{Fig1}d and, RDFs for other trajectories are in Figure S4). 
Upon urea-induced destabilization, the $\mathrm{g_u}(r|\mathrm{B})$ increases with a pronounced peak at 4 ${\rm \AA}$, indicating that urea molecules make direct interactions with the disrupted bases, while $\mathrm{g_w}(r|\mathrm{B})$ shows only a minor increase. 
The increase of $\mathrm{g_w}(r|\mathrm{B})$ is simply a consequence of enhanced exposure of the denatured bases to the solvent environment. 
The sharp peak of $\mathrm{g_w}(r|\mathrm{P})$ at 2 \AA\ (Figure \ref{Fig1}d) is due to formation of hydrogen bonds between water hydrogen and phosphate oxygens (see Figure S5).
Compared with $\mathrm{g_u}(r|\mathrm{B})$, changes in $\mathrm{g_u}(r|\mathrm{S,P,Q})$ are all minor, suggesting that major change in urea distribution upon denaturation occurs around nucleobases, which lends support to our earlier proposal that urea acts as a surrogate base. 
\\

\noindent \textit{Changes in torsion angles: }
Signature of structural destabilization due to urea is reflected in the changes in torsion angles as well. 
The distributions of seven torsion angles of phosphodiester backbone, calculated for the first and last 10 ns of the trajectories at 8 M urea, reveal that the most distinct changes are in the $\zeta$ value defined in terms of O5'-P-O3'-C3' atoms (Figure 1e and Figure S6a).  
Consistent with our previous study \cite{Priyakumar09JACS}, in aqueous urea solutions the major peak of $P(\zeta)$ at $300^{\circ}$ shifts to $270^{\circ}$, and a new minor peak at $30^{\circ}$ is formed.  However, we also found that     
$P(\zeta)$ contributed by the tertiary pairs differs greatly from the $P(\zeta)$ associated with the secondary pairs (\ref{Fig1}e and Figure S6b). 
While $\zeta$-values of bases in the stem region are mainly distributed around 300$^o$, G13 and U24 have distinct $\zeta$ values because they stack with preQ$_1$ to form a sharp turn toward U28 via A25. 
The $\zeta$ values for the bases in the A-rich tail vary from base to base and exhibit dramatic changes upon disruption of the base pairs in the sharp turn (U24-U26) and the L1-loop (Figure S6b). 
\\

\noindent{\bf Denaturation of secondary and tertiary motifs of the RS.} 
In aqueous urea, base pairs of RNA exhibit rearrangement as well as disruption. 
The details of structural changes in RS are determined by interplay of interactions associated with water, urea, and the degree to which RNA bases are exposed to solvent molecules. 
Differences in sequence and local environment lead to heterogeneous dynamics of urea-induced destabilization in each base pair. \\

\begin{figure}
\centerline{%
\includegraphics[scale=0.36]{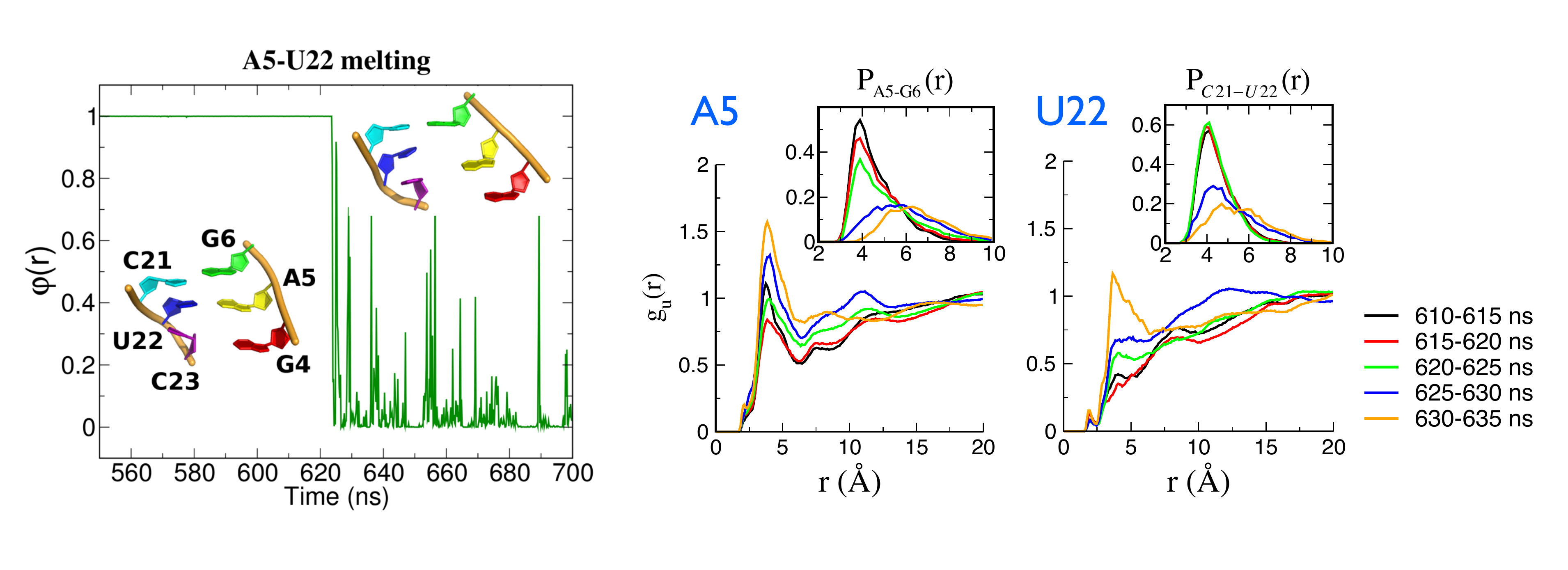}%
}%
\caption{Denaturation of local structures. Destabilization of A5-U22 base pair in 8 M urea quantified using by $\varphi(r)$ (Eq. \ref{eqn:logistic}) and corresponding changes in RDF of urea during the process of base pair disruption. 
Pair distribution functions (insets) show that urea can solvate disrupted bases only when stacking between neighboring bases are removed.
}
\label{Fig2}
\end{figure}

\noindent {\it Removal of stack from neighboring bases is critical for base-pair disruption}:  
%Disruption of base pair at the end of P1 helix stack provides urea with better accessibility to the base pairs inside the stack. 
For a disrupted base to have maximal exposure to the urea environment,  
the base should be released from the influence of the nonpolar stacking interaction with the neighboring bases. 
The disruption dynamics of A5-U22 base pair that lies at the center of P1 helix, followed by the disruption of G4-C23 pair, shows this process clearly. 
The increases in urea population around A5 and U22 occurs upon breaking of A5-U22 base pair, at 620-630 ns (Figure 2). 
The distribution of A5 relative to the neighboring base G6, quantified by the pair distance distribution ($P_{\rm A5-G6}(r)$) shifts from 4 \AA\ to 6 \AA\ and broadens with time (see inset in Figure \ref{Fig2}), confirming that A5 is released from the stack with G6.  
A similar pattern of time evolution is seen for U22 in $P_{\rm C21-U22}(r)$ (Figure \ref{Fig2}). \\

\begin{figure}
\centerline{%
\includegraphics[scale=0.31]{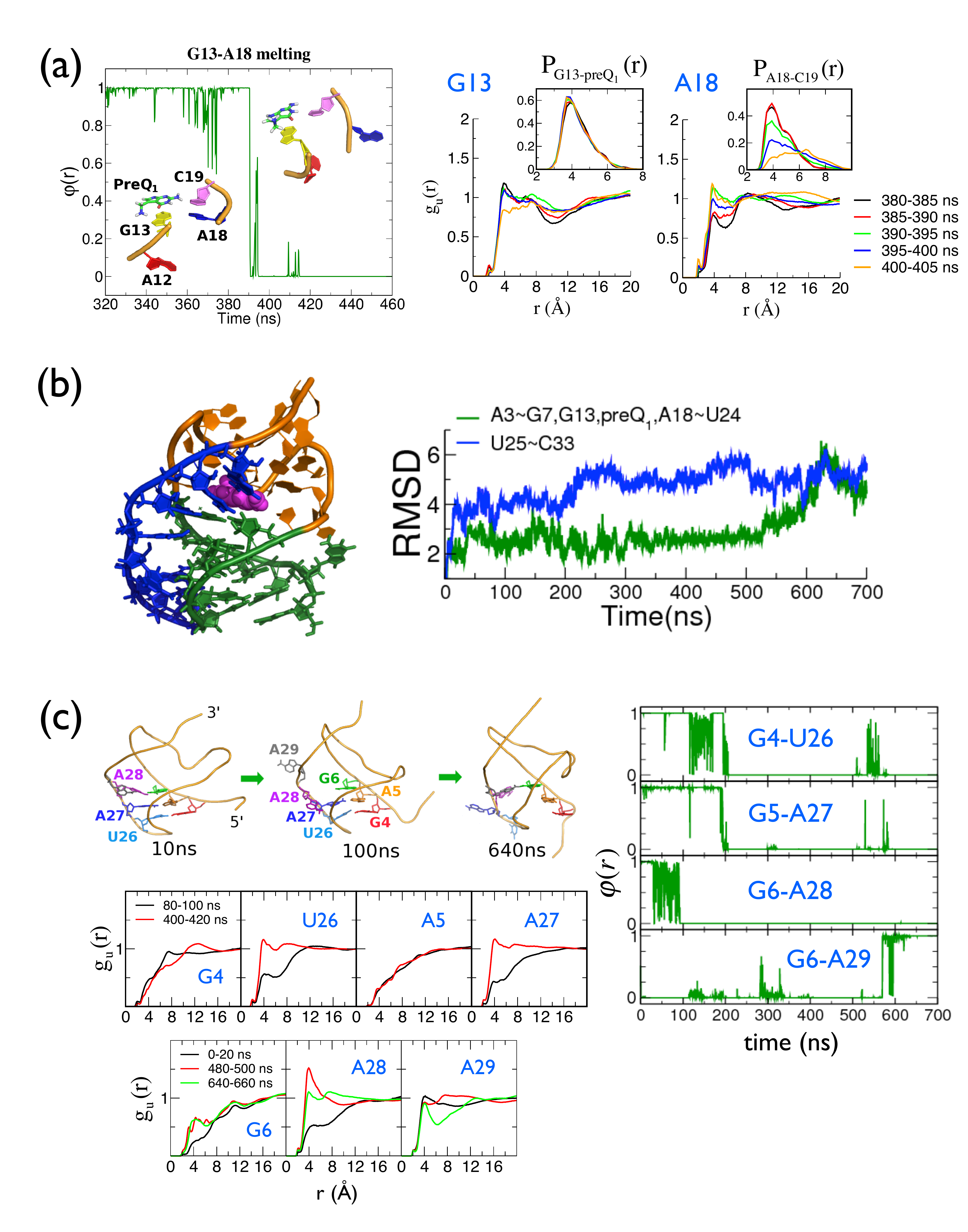}%
}%
\caption{Denaturation of local structures. 
{\bf (a)} Dynamics of disruption of G13-A18 base pair in 8 M urea quantified by the order parameter $\varphi(r)$ (Eq. \ref{eqn:logistic}) and the  corresponding changes of  RDF for urea during the process of base pair disruption. Pair distribution functions (insets) show that urea can solvate disrupted bases only when stacking between neighboring bases are removed.
{\bf (b)} RMSD of stem+loop region (A3$\sim$U24) and tertiary structure region (U25$\sim$A33) reflect the difference in flexibility between the two region. 
%RS figure at left side demonstrates that two stable adjacent G-C pairs (G6-C21, G7-C20) and Ca$^{2+}$ ions stabilize stem+loop region, leading to low fluctuations in RMSD.
{\bf (c)} Disruption and rearrangement of tertiary base pairs of the RS in 8 M urea. Conformational changes of G4, A5, G6 and A26, A27, A28, A29 are illustrated in the snapshots. $\varphi(r)$ (Eq. \ref{eqn:logistic}) for each base pair shows cooperative disruption of G4-U26 and A5-A27 base pairs and  rearrangement of G6-A28 $\rightarrow $ G6-A29 at 8 M urea. G6-A28 pair, disrupted at 100 ns, is replaced by G6-A29 at 620 ns. $\mathrm{g_u}(r)$ around each base are shown.}
\label{Fig3}
\end{figure}

\noindent\textit{Urea destabilizes tertiary structures}: 
Urea induced dynamics of base pairs in tertiary motifs is considerably more complex than those found in  secondary motifs. 
For example, the non-WC G13-A18 pair, located in the L2 loop with G13 being coordinated to preQ$_{1}$, 
undergoes a more complicated disruption process than the A5-U22 pair (Figure \ref{Fig3}a). 
Examination of the snapshots of structures from simulation (Figure \ref{Fig3}a) shows that  
A18 is fully exposed to the solvent environment while G13 is in complex with preQ$_1$ even after the G13-A18 interaction is disrupted. 
As a result, urea population around G13 remains unchanged while A18 attracts urea molecules.  
There is no significant change in $P_{\rm G13-preQ_{1}}(r)$ even after the disruption of G13-A18 pair, which indicates that G13 maintains the stacking interaction with preQ$_1$.  

Base pairs in tertiary motifs are less stable than those in secondary structures. 
As shown by the RMSD calculated separately for secondary and pseudoknot stacks (Figure \ref{Fig3}b), base pairs at the stem region of secondary structure remain compact, while tertiary structure, prone to be exposed to the solvent environment, expands readily.
The cooperative disruption of Hoogsteen base pairs (G4-U26 and A5-A27) responsible for forming pseudoknot stacks, is followed by base pair rearrangement of G6-A28 to G6-A29. 
The logistic function $\varphi(r)$ in Figure \ref{Fig3}c quantifies how this rearrangement occurred in 600 ns after the base pair disruption. 
The $\mathrm{g_u}(r)$ (Figure \ref{Fig3}c) and $\mathrm{g_w}(r)$ (see Figure S7) summarize the changes in solvent environment around tertiary motifs before and after the denaturations of  G4-U26, A5-A27 pairs and G6-A28$\rightarrow$G6-A29 rearrangement. 
Both $\mathrm{g_u}(r)$ and $\mathrm{g_w}(r)$ around G4 and A5 remain unchanged during the denaturation process of G4-U26 and A5-A27 because the base pairs G4-C23 and A5-U22 that stabilize the P1 helix are still intact, making G4 and A5 inaccessible to the solvent (Figure \ref{Fig3}c). 
In contrast, populations of urea and water around U26 and A27 show noticeable increases (Figure \ref{Fig3}c and Figure S10). 

Similar to G4 and A5, 
the disruption of Hoogsteen pair G6-A28 does not significantly affect the distributions of urea and water around G6 (Figure \ref{Fig3}c) because the Watson-Crick base pairing of G6 with C21 in the stem region is still intact (Figure \ref{Fig1}). 
The disruption of G6-A28 increases solvent density around A28 only. 
In contrast, urea density around A29 decreases due to the formation of G6-A29 interaction.
Taken together, these results show that the degree of urea solvation strongly depends on the degree of interbase interactions. 
Most significantly, the stacking interactions between neighboring bases are the key determinant for urea's accessibility and subsequent disruption of the structure.\\

\begin{figure}
\centerline{%
\includegraphics[scale=0.25]{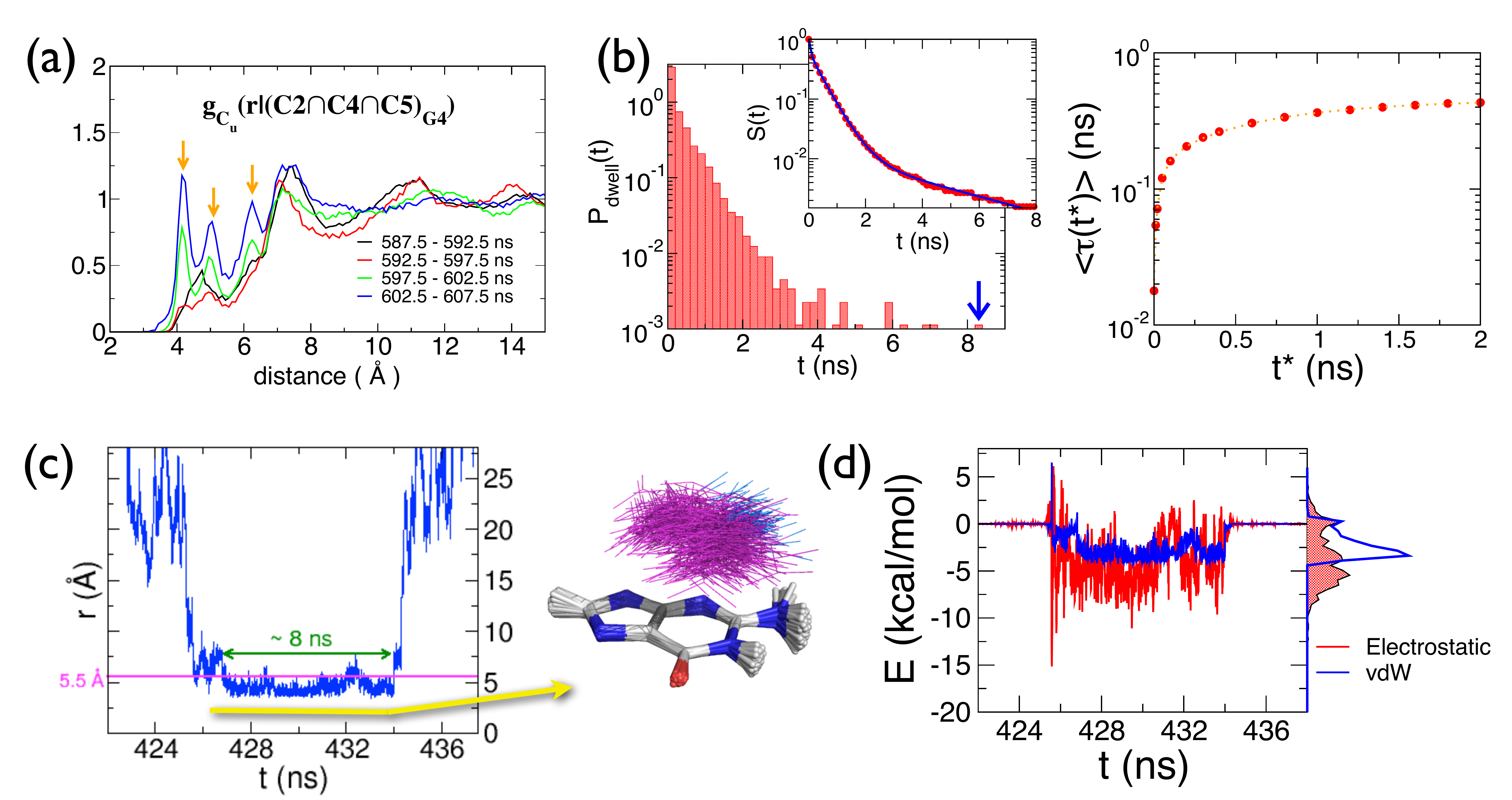}%
}%
\caption{Dynamics of stacking between urea and nucleobase. 
{\bf (a)} Development of $\mathrm{g_{C_u}(r)}$ between urea carbon and group of carbon atoms in purine ring of G4 during disruption of 
G4-C23 base pair. 
The peaks at 4.2 \AA, 5 \AA, and 6.3 \AA\ (marked with yellow arrows) reflect formation of stacking and hydrogen bonding interaction between urea and G4 base as a function of time. 
{\bf (b)} (Left) Dwell time distribution $P_{dwell}(t)$ and survival probability $S(t)$ for stacking interaction with $t^*=1$ ns. 
$S(t)$ can be fit to tri-exponential function: $S(t)=\sum_{i=1}^3\phi \exp{(-t/\tau_i)}$ with 
$\phi_1=0.591$, $\tau_1=0.484$ ns, 
$\phi_2=0.395$, $\tau_2=0.058$ ns, 
$\phi_3=0.014$, $\tau_3=3.42$ ns, which give the average life time of stack is $\langle\tau\rangle=\int_0^{\infty}dt S(t)=0.365$ ns. 
One of the longest dwells in the simulation trajectories is $t\approx 8.2$ ns, which is marked with an arrow in $P_{dwell}(t)$. (Right) $t^*$-dependent average lifetime of stack is computed using $\langle \tau(t^*)\rangle=\int^{\infty}_0dtS(t;t^*)$. 
The lifetime of geminate pair of urea-base stack is $\langle\tau(t^*\rightarrow \infty)\rangle=0.46$ ns. 
{\bf (c)} The stack forming trajectory corresponding to $\approx 8$ ns dwell time in (b) is shown in the bottom panel. Note that transient disruption ($r>5.5$ \AA) less than $t^*=1$ ns is still considered as a part of dwell. 
Structural ensemble of urea molecule stacked on guanine base. 
Urea oxygen is colored in blue.  
{\bf (d)} Stacking energy decomposed into electrostatic and vdW interactions. 
}
\label{Fig4}
\end{figure}

\noindent{\bf Stacking and hydrogen bonding dynamics of urea with nucleobases.}
Although increase of $\mathrm{g_u}(r|\mathrm{B})$ suggests that urea stabilizes denatured nucleobases (Figures \ref{Fig2}, \ref{Fig3}), the details of interaction between urea and denatured nucleobase are not well captured by the static $\mathrm{g_u}(r|\mathrm{B})$ alone.
A careful inspection of the simulation trajectories provide insights into urea configurations that form stacks and hydrogen bonds with bases.  
In order to visualize the presence of such configurations, it is important to define RDFs in atomistic level. 
For instance, to identify a configuration of urea stacked on a guanine base, we calculated the RDF between urea carbon ($C_u$) and the three carbons of guanine base (C2, C4, C5), i.e., ${\rm g}_{C_u}(r|(C2\cap C4\cap C5)_{G4})$ where $r$ is the minimum distance from $C_u$ to any of $C2$, $C4$, and $C5$ in $G4$ (see Figure \ref{Fig4}a). 
In the case of hydrogen bond, RDF is straightforwardly defined either between urea oxygen and base hydrogen or between urea hydrogen and base oxygen (or nitrogen) (Figs 6b, S9, S10b).

\begin{figure}
\centerline{%
\includegraphics[scale=0.25]{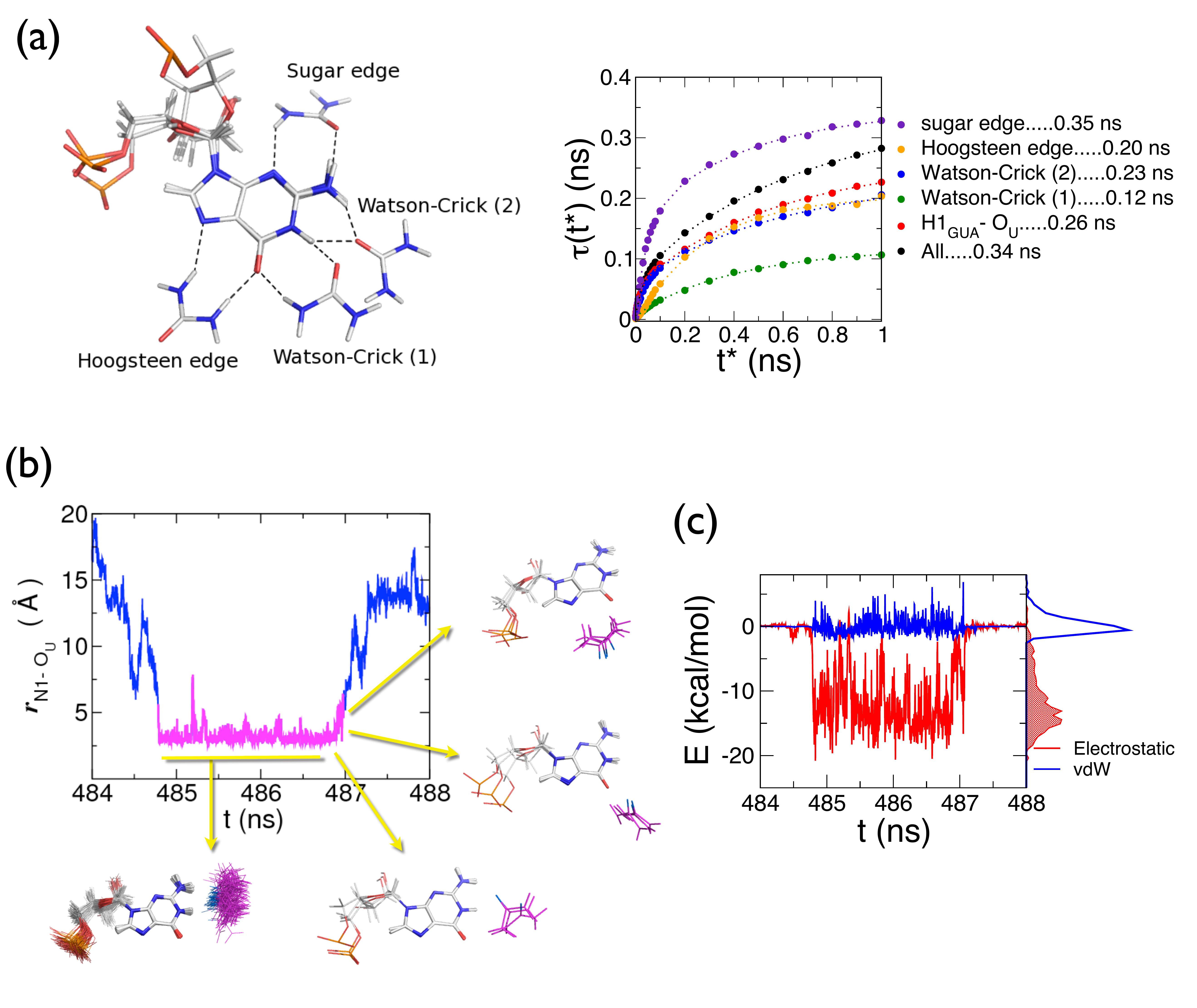}
}%
\caption{Dynamics of hydrogen bonding between urea and nucleobase. 
{\bf (a)} $t^*$-dependent average lifetimes of five different types of hydrogen bonds made between urea and base. 
$\tau(t^*\rightarrow\infty)$ values for different bond types are given in the legend. 
Configuration of Watson-Crick (1), (2), Hoogsteen, sugar edge hydrogen bonds are shown on the right. ``All" denotes the union of five types of hydrogen bonding interactions.
{\bf (b)} The part of time trace colored in purple corresponds to the time interval in which urea-guanine hydrogen bond is made. 
Note that the urea can have various hydrogen bonded configurations with base, migrating around the nucleobase while still satisfying the criterion for hydrogen bond. 
{\bf (c)} The energetic contribution of H-bonds decomposed into electrostatic and vdW interactions suggests that electrostatic interaction is the dominant driving force for hydrogen bond formation.     
} 
\label{Fig5}
\end{figure}

The static RDF does not provide the duration of the interactions between base and an individual urea molecule. 
In solution more than one urea molecules can interact with a base at a given time.   
Even if a particular urea-base pair lasts only transiently, exchange with other urea can effectively prolong the life time of these interactions. 
Hence, the distribution of urea around a particular base would be constant if the time scale for averaging is sufficiently long. 
What is the lifetime of urea-base interaction?    
To answer this question, we computed the dwell time distributions for both hydrogen bonding and stacking interactions (Figures \ref{Fig4} and \ref{Fig5}) defined as  
\begin{equation}
P_{dwell}(t;t^*) = \dfrac{1}{N_{dwell}} \sum_{i=1}^{N_{dwell}} \delta [\tau_i(t^*)-t]. 
\label{eqn:Dwell} 
\end{equation}
Because the stacked or hydrogen-bonded configuration, defined based on certain geometrical criteria, is disrupted but   reforms transiently, we take the transient disruption time $t^*$ into account in defining the dwell time. This allows us to ignore the transient disruption and reassociation processes whose time lapse is shorter than $t^*$.   
In the Eq. \ref{eqn:Dwell}, $\tau_i(t^*)$ is the time interval of the $i$-th dwell for stacked (or hydrogen bonded) configuration between urea and base; 
$\delta[\tau_i(t^*)-t]=1$ if $t<\tau_i(t^*)<t+dt$ where $dt$ is the size of time window used to calculate the histogram, otherwise $\delta[\tau_i(t^*)-t]=0$;  
$N_{dwell}$ is the number of dwells observed over the entire time traces with the subscript $i$ denoting the index of $i$th dwell. 
Using the dwell time distribution for a given $t^*$, $P_{dwell}(t,t^*)$, we can also calculate the survival probability as
$S(t;t^*) = 1 - \int_{0}^{t} P_{dwell}(\tau;t^*) d\tau$. Thus, the average dwell time for a given $t^*$ is calculated using $\langle\tau(t^*)\rangle=\int^{\infty}_0S(t;t^*)$; and the lifetime of a ``geminate" pair between urea and base can be calculated by setting $t^*\rightarrow \infty$ (in practice the upper limit corresponds to duration of the simulation). 
Below we employ these definitions to evaluate the transient ($t^*=0$) and geminate pair ($t^*\rightarrow\infty$) lifetime of stacked and hydrogen bonded urea-base pairs. 
\\

\begin{figure}
\centerline{%
\includegraphics[scale=0.22]{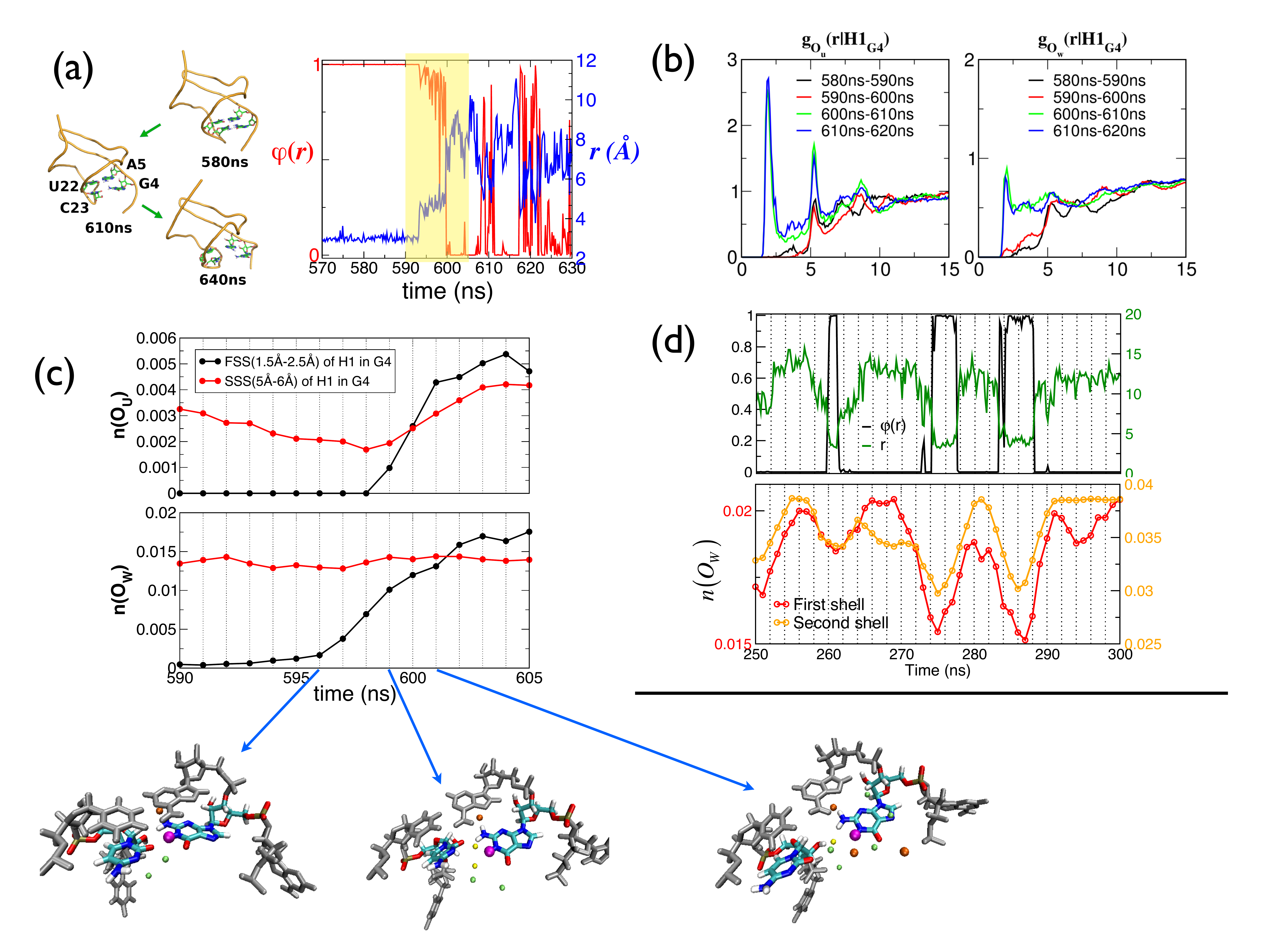}%
}%
\caption{Interplay of dynamics between urea and water molecules in the process of denaturation. 
{\bf (a)} Sequential disruption of secondary structure at stem region. 
The region highlighted in yellow is when G4-C23 base pair disruption occurs. 
{\bf (b)} Change in $\mathrm{g_u}(r)$ and $\mathrm{g_w}(r)$ around H1 hydrogen atom participating in G4-C23 base pair interactions. 
{\bf (c)} Time evolution of density of oxygens of urea ($n(\rm{O_u})$) and water ($n(\rm{O_w})$) in the first solvation shell (FSS) and second solvation shell (SSS) of H1 atom of G4 during G4-C23 base pair disruption. Density of oxygen atoms was calculated using 
$n(O)=N_{shell}/V_{shell}$ where $N_{shell}=4\pi\rho\int_{r_{min}}^{r_{max}} g(r)r^{2} dr$, $V_{shell}=4 \pi {r_{c}}^2 \sigma_{shell}$, $\rho$ is a bulk density of solvent, $g(r)$ is a radial distribution function, [ $r_{min}$, $r_{max}$ ] specify the location of a shell, $r_{c}=(r_{max}+r_{min})/2$, and $\sigma_{shell}=r_{max}-r_{min}$. Three snapshots visualize the permeation dynamics of water molecules in the FSS (yellow spheres) and SSS (green spheres), which fill the space between base pair before a urea oxygen (red sphere) form hydrogen bond with H1$_{G4}$ (magenta sphere) (see also SI movie 5). 
{\bf (d)} Hydration dynamics for reversible disruption of A3-U24 base pair for urea-free simulation trajectory in the absence of both preQ$_1$ and Ca$^{2+}$. Shown are the $\varphi(r)$ (Eq. \ref{eqn:logistic}) for N1$_{A3}$-N3$_{U24}$ distance and time dependent $n(O_\mathrm{w})$ in FSS and SSS. 
$n(O_\mathrm{w})$ and $\varphi(r)$ are anti-correlated, which implies that disruption of base pair interaction is water mediated.  
}
\label{Fig6}
\end{figure}

\noindent {\it Stacking dynamics and energetics of urea with nucleobase}:
When G4 is released from its Hoogsteen partner C23 in 8 M urea at $t\approx 600$ ns (see Figures \ref{Fig1}c and \ref{Fig6}a), 
the RDF of urea carbon around C2, C4, C5 atoms of G4, i.e., $\mathrm{g}_{C_u}(r|(C2\cap C4\cap C5)_{G4})$ (Figure \ref{Fig4}a) develops three peaks, which reflects the configurations of urea that hovers above the G4 base ring. 
To analyze the dwell time kinetics of urea-base stack from time trajectories, we set a distance criterion for urea-base stack: max$\{r_{\mathrm {C_{u}-C2_{G4}}},r_{\mathrm {C_{u}-C4_{G4}}},r_{\mathrm {C_{u}-C5_{G4}}}\}< 5.5 {\rm \AA}$. 
The dwell time distribution and survival probability calculated for various $t^*$ values reveal that although lifetime of stacking for $t^*=0$ is less than 0.02 ns, the stack lifetime converges to 0.46 ns when $t^*\rightarrow\infty$ (Figure \ref{Fig4}b). 
$P_{dwell}(t;t^*)$ and $S(t;t^*)$ of stacking interaction for $t^*=1$ ns are shown on the left panel of Figure \ref{Fig4}b. 
The event marked by the blue arrow in the $P_{dwell}(t)$ is due to a dynamic time trace of urea-base stack with very long dwell time, the actual time trace of which is shown in Figure \ref{Fig4}c in terms of urea-base distance as well. 
The trajectory in Figure \ref{Fig4}c shows how a urea molecule approaches the base, maintains stacking interaction for about 8 ns, and subsequently diffuses away (see the SI movie 1).      
The conformation of urea structures that satisfy the above-mentioned distance criterion for stack lies parallel to the purine ring of a guanine base.
% with urea oxygen pointing opposite to the O6$_{G4}$ and close to two hydrogens attached to the N2$_{G4}$. 
It is important to note that the space between urea-base stack is ``dry", devoid of any water molecule (Figure \ref{Fig4}c). 
Despite the lack of chemical similarity to purine or pyrimidine bases, it is remarkable that urea acts as a base when interacting with guanine. 
%The stacked configurations suggest that the first two peaks in Fig.\ref{Fig4}a are due to stacking interaction, but the third peak is possibly from hydrogen bonds.

The non-covalent energy associated with urea-base stacking can be decomposed into electrostatic and van der Waals (vdW) interactions.  
For the urea stacking on the guanine base the electrostatic interactions make a greater energetic contribution ($\sim -5$ kcal/mol) than the vdW interaction ($\sim -3$ kcal/mol) (Figure \ref{Fig4}d). 
The larger energetic contribution from electrostatic interaction is due to the attractive interaction between the O6$_{G4}$ and urea hydrogens, and between urea oxygen and base hydrogens. 
It is worth emphasizing that the ensemble of urea-guanine stack in Figure \ref{Fig4}c displays the antiparallel dipole-dipole configurations, in which the urea oxygen faces away from the O6$_{G4}$ to minimize repulsive interactions with O6$_{G4}$ and to maximize attractive interaction with hydrogens of N1$_{G4}$ and N2$_{G4}$. 
For urea-adenine stack, the antiparallel dipole-dipole interaction is not as stable as urea-guanine stack since the carbonyl group of guanine base is replaced by amine group in adenine, which renders electrostatic contribution to urea-adenine stack essentially zero. 
However, the urea-adenine stack is stabilized by nearly the same amount of vdW interaction ($\sim -3.5$ kcal/mol) (Figure S8a). 

Urea stacking with pyrimidine bases is not as stable as purine (guanine, adenine) ring because of the steric hindrance or interaction with phosphate and ribose groups. 
%The steric hindrance from ribose group always prevents urea from stacking on pyrimidine ring. 
As shown in SI movie 2, 
urea spends most of the time interacting with phosphate group or ribose, which makes urea-cytosine base stacking incomplete. 
%electrostatic interaction between phosphate group or ribose ring of backbone can only allow urea to form a loose stacking with cytosine base. 
For uracil, with the help from carbonyl group located at position 4, urea can form more stable vdW interaction (-2.7 kcal/mol) than cytosine (-1.8 kcal/mol). 
%Urea stacking with cytosine base is loose 
%than with uracil base to avoid repulsive electrostatic interaction which makes electrostatic interaction between urea and cytosine base more stable than urea-uracil electrostatic interaction 
(Figure S8c and SI movie 3).   

Taken together these results show that stacking interactions of urea with the bases have greater effect on purines than pyrimidine bases. 
In addition, it is of note that although the energetic contribution of the vdW energy due to urea-base stacking is smaller than the electrostatic interactions (Figure \ref{Fig4}d and Figure S8), the nonzero value of vdW energy is germane to the urea-base stacks (see below or Figure \ref{Fig5}c to compare the vdW energy in urea-base stack with that of urea-base H-bond); thus, one can use vdW energy between urea molecules and nucleobases to report on the progress of urea-induced denaturation of nucleic acids.  
\\

\noindent {\it Hydrogen bond dynamics and associated energetics of urea with nucleobase}:
Analyses of the trajectories show that there are a few distinct types of hydrogen bonds between urea and nucleobase, which include Watson-Crick, Hoogsteen, and sugar edge hydrogen bonds.  
We defined the formation of hydrogen bond by using distance and angle constraints, $r_{\mathrm D - A}<$ 3.5$\rm \AA$, $|\theta_{\mathrm D-H \cdots A}-180^{\circ}|<45^\circ$, $|\varphi_{\mathrm H_U \cdots X1-X2-X3}|<45^\circ$, where D and A are hydrogen donor and acceptor, respectively, and $X1$, $X2$, $X3$ are the covalently linked heavy atoms in the base ring, which form a dihedral angle of $\pm 45^\circ$ with urea hydrogen. 
We relaxed the bond and dihedral angle cutoffs from the usual single hydrogen bond cutoff value of $30^\circ$ (Ref. \cite{Luzar96PRL}) to $45^\circ$ because $30^\circ$ is too stringent to accommodate the geometry of a urea molecule that interacts with base using multiple hydrogen bonds (see Figure \ref{Fig5}a). 

Interestingly, bond lifetimes vary greatly depending on the context and type. 
%and not all of them contribute significantly to overall urea-purine base hydrogen bonding. 
When $t^*$ is small (less than 5 ps) the lifetime of $\mathrm {O_{U} - H1_{G}}$ hydrogen bond is the longest, while as $t^{*} \rightarrow \infty$ the sugar edge hydrogen bond shows the longest lifetime, implying that $\mathrm {O_u - H1_{G}}$ hydrogen bond is energetically more stable while sugar edge hydrogen bonding hold urea longer. 
The lifetimes of hydrogen bondings increase and converge to the values of (0.1$-$0.4) ns as $t^{*} \rightarrow \infty$. 

Similar to the time trace associated with stacking dynamics in Figure \ref{Fig4}c, the time trace of the distance between 
$\mathrm{O_u}$ and $\mathrm{N1_G}$ in Figure \ref{Fig5}b displays remarkably complex dynamics.  
The part of the trace colored in magenta corresponds to the instance urea-guanine hydrogen bond is formed. 
Snapshots in Figure \ref{Fig5}b show that urea migrates around guanine base by forming different types of hydrogen bonds. 
At $t\approx$ 485 ns, $\mathrm{O_u}$ mainly forms hydrogen bond with hydrogen atoms in N1 and N2 of guanine. 
At $t\approx$ 487 ns, $\mathrm{O_u}$ migrates to O6 to form O6$-$urea hydrogen interaction, followed by Hoogsteen type interaction, and finally diffuses away from the nucleobase (see also the SI movie 4). 
When the H-bond between urea and base is decomposed into electrostatic and vdW contributions, it is clear that the dominant energy source of H-bond is electrostatic; the vdW interaction makes practically no contribution (Figure \ref{Fig5}c). 
\\

\noindent{\bf Dynamic role of water molecules in the process of urea-induced denaturation.}
For urea-induced denaturation to occur, both thermodynamic and kinetic requirements should be met so that urea can form a favorable interaction with the denatured bases; and the intrusion of urea into the narrow space between base-base pairing should occur. 
Due to the smaller size, compared to urea, water molecules can diffuse more easily into various regions of the nucleobases, while urea cannot easily access the space between the stacked bases. 
%which makes $\mathrm{g_u}(r)$ before and after the melting more distinct (Fig.\ref{Fig1}d).
It is worth noting that in our simulations in 8 M urea, the diffusion coefficient calculated using mean-square displacement of water  (2.3$\times 10^{-5}$ $cm^{2}/s$) is twice greater than that of urea ($10^{-5}$ $cm^2/s$).
 \\

\noindent{\it Microstructure of solvent environment around nucleobases }:
To investigate the solvation process in further detail, we calculated the RDFs of urea and water oxygens around $\mathrm{H1_{G4}}$ ($\mathrm{g_{O_u}}(r|\mathrm{H1_{G4}})$, $\mathrm{g_{O_w}}(r|\mathrm{H1_{G4}})$),
so that the details of the H-bond disruption associated with G4-C23 pair (Figure \ref{Fig6}) could be easily detected.  
During the destabilization process of the stem region that occurs from the ends of the stack (disruption of G4-C23 pair, followed by A5-U22) (Figure \ref{Fig6}a), two sharp peaks develop in $\mathrm{g_{O_u}}(r|\mathrm{H1_{G4}})$ at 2 \AA\ and 5$\rm \AA$.
The first peak at 2 \AA\ is due to the hydrogen bond between urea oxygen and H1 of the guanine base.   
The second peak at 5$\rm \AA$ that appears in the disrupted purine base, but not in the  pyrimidine base (see Figure S9), is in fact contributed by the stack between urea and purine base. 
This finding is in sharp contrast to the relatively structureless $\mathrm{g_{O_w}}(r|\mathrm{H1_{G4}})$ (Figure \ref{Fig6}b). 
\\

\noindent \textit{Water permeation precedes urea-base interactions }:
A careful inspection of RDFs in Figure \ref{Fig6}b provides further insight into the dynamics associated with urea and water.   
At the early stage of destabilization (590$-$600 ns), urea density at $(3-5)$ \AA\ near the second solvation shell (SSS) (peak at 5$\rm \AA$) of $\mathrm{g_{O_u}}(r|\mathrm{H1_{G4}})$ decreases slightly while water density increases in both the FSS and SSS. 
Increase of water in the region around the FSS is due to  the replacement of inter base hydrogen bonds by water.  
When G4-C23 base pair is fully disrupted (600-620 ns), there is a marked increase in urea population in the FSS and SSS, 
which correspond to formation of  hydrogen bond and stacking interactions, respectively. 
To assess the urea and water dynamics during base pair disruption quantitatively, we calculate  
the time evolution of the number density of urea oxygen ($n(\mathrm{O_u})$) and water oxygen ($n(\mathrm{O_w})$) in the FSS and SSS of H1 atom in G4 base (Figure \ref{Fig6}c).  
Note that motions of water, urea, and nucleobase are highly dynamic and undergo substantial fluctuations even in the time interval as short as 1 ns. To see the overall changes in the number density of water and urea, we filtered the fluctuations in data by taking the running average over 5 ns. 
As is clearly shown in Figure \ref{Fig6}c, destabilization of the base pair occurs in the following three steps: 
(1) In the initial stage ($590\sim 598$ ns), increase of $n(\mathrm{O_w})$ in the FSS is accompanied by the decrease of $n(\mathrm{O_u})$ in the SSS, which indicates that water permeates into the FSS 
by expelling urea away from the SSS. This process breaks the hydrogen bonds between base pairs. 
It is important to note that at $t\approx 598$ ns when water is present in FSS the density of urea is effectively zero (Figure \ref{Fig6}c). 
(2) In the disruption stage ($598\sim 602$ ns), the numbers of both urea and water molecules increase drastically in the FSS and SSS. The disrupted bases are stabilized by forming multiple hydrogen bonds and stacks with urea.  
(3) In the late stage ($602$ ns $\sim$), the water population in the FSS and SSS stops increasing, but urea increases further to fully solvate the bases. 
The mechanism of cooperative action between urea and water, revealed in the disruption of secondary structure, also holds in the disruption of tertiary structure (see SI Text and Figure S10a). 
Among the steps (1) $-$ (3), of particular note is the initial infiltration of water molecules into the dry and narrow space between base pair (step (1)), which is an important prerequisite for urea to interact with nucleobases (see SI movie 5 that visualizes water molecules in the FSS (yellow) and SSS (green) in the time interval of 596 ns $-$ 601 ns). 
Only after destabilization of base pair interaction by penetration of  water can urea solvate the nucleobases. 

To verify that base pair disruption is indeed initiated by water permeation, we additionally examined the hydration dynamics around A3-U24 base pair from a urea-free simulation in the absence of preQ$_1$, and Ca$^{2+}$ (Figure \ref{Fig6}d). 
As expected, the fluctuation of base pair distance is correlated with water densities in the FSS and SSS.    
It is remarkable that when the time series of water density around A3 and breathing dynamics of base pair are compared, the change of water density {\it always} precedes the change in base pair distance. 
The increase in water population begins to slow down only after the disruption of base pair. 
Conversely, the water population stops decreasing after base pair formation is completed. 
Thus, it follows that the dynamics of water permeation gives rise to the breathing dynamics of base pairs. 
Urea can stabilize this transiently unbound base pair by forming stacks or multiple hydrogen bonds, which results in denaturation of the base pair.  
\\

\section{Concluding Remarks}
Unlike proteins, base pairs, a major building block of structured nucleic acids, are always deeply buried inside the major and minor grooves of folded RNA molecules; thus there is no water solvation shell that can directly interact with the nucleobases. 
Displacement of  water in the first solvent shell  by urea, which was suggested as an early stage dynamics for urea-induced protein denaturation \cite{hua2008PNAS}, cannot account for urea-induced RNA denaturation. 
In sharp contrast, our simulations highlight the early penetration of water in the narrow space between the base pairs or base stack as a critical kinetic step that {\it ought to precede} before urea interacts with the bases. 
Since the phosphodiester backbone does not particularly attract urea molecules, and urea is too big to fit between the nucleobases, the enhancement of structural fluctuations due to water-base interaction is critical for urea to access and interact with nucleobases. 
Thermodynamic stabilization of disrupted nucleobases by urea's multiple hydrogen bonds and stacking interactions follows afterward. Based on these findings we propose that the kinetic sequence of destabilization of folded RNA should follow native state $\rightarrow$ wet state $\rightarrow$ unfolded state.  
Because the lifetime of germinate urea-nucleobase pairs is short ($< \mathcal{O}(1)$  ns) (Figs. 4b and 5a) compared with the breathing time of base pairs ($\sim 10$  ns) (Figure 6d), the sequence of events involving urea urea around nucleobases is highly dynamic.

The high polarity and size comparable to water are the unique properties of urea, which enable urea to have high solubility in water without perturbing the structure of water  \cite{Wallqvist97JACS}, and interact with various components of nucleic acids by exploiting the ability to engage in multitude of hydrogen bonds. 
Besides the urea's hydrogen bonding with nucleobases, we highlighted the ability of urea to stabilize disrupted nucleobases via stacking interactions.
%It is noteworthy that although methyl derivatives of urea may increase the molecule's vdW interaction with nucleobases \cite{Herskovits63Biochemistry}, such methyl functionalization inevitably of molecule. 
Our study shows that the ability of urea to engage in both stacking and hydrogen bond interactions makes it an effective denaturant of nucleic acids. 
%To conclude, aqueous urea is one of the most effective denaturant of both proteins and nucleic acids. 

\section{Methods}
\paragraph{Riboswitch model}
The \textit{B. subtilis} preQ$_{1}$-riboswitch aptamer is the only one whose solution and crystal structures have been determined \cite{Zhang11JACS}. 
Although the solution and the crystal structure are nearly identical, there are some differences. 
The crystal structure has several visible Ca$^{2+}$, especially two Ca$^{2+}$ ions in L1-P2 turn while Ca$^{2+}$ is not present in solution structure. 
First, it has been shown that Ca$^{2+}$ ions stabilize the L2 loop. 
Secondly, L2 residues are well-defined in the solution structure while they are not resolved in the crystal structure. 
In our simulations we included two Ca$^{2+}$ ions in L1-P2 turn from crystal structure into the solution structure (PDB id: 2L1V).   
   
%\subsection{Simulation procedure\\}
\paragraph{Simulation details}
PreQ$_{1}$-riboswitch is solvated in a 60 \AA\ $\times$ 60 \AA\ $\times$60 \AA\ box containing 6267 TIP3P water. For 8 M urea simulation, 1052 urea molecules were randomly distributed by replacing water molecules in the box. After a short duration when energy minimization was carried out to remove clashes among urea molecules, 51 Na$^{+}$ and 20 Cl$^{-}$ ions, which amount to $\approx$ 150 mM salt concentration in the bulk, were randomly placed. 
%We performed energy minimization using 1fs time step for 3000 steps with the constraint on PreQ$_{1}$-riboswitch coordinates. 
At the equilibration stage we gradually heated the system from 0K to 310 K using 2 fs time step, % for 620 ps with constraint, 
and simulated for an additional 100 ps in the NPT ensemble and an additional 100 ps NVT ensemble without constraint to ensure pre-equilibration at 310 K. We further equilibrated the urea-riboswitch system for 10 ns by using the NPT ensemble at 310 K and 1 atm. 
Theories and experiments \cite{ThirumARPC01,Koculi07JACS,moghaddam09JMB} show that counterions play a key role in the folding of RNA. The distribution of monovalent counterions (Na$^+$) on the surface of RS after equilibration is shown in Figure S12. High density of negative charge is formed at core regions of riboswitch structure where base pair interactions are made, which is indicated by the value of electrostatic potential from a solution of nonlinear Poisson Boltzmann equation. It is worth noting that major population of counterions, calculated from time trajectories from MD simulations, are in the core regions predicted by the nonlinear Poisson Boltzmann equation to have high negative electrostatic potential. 
We performed simulations by using NAMD with CHARMM force field for riboswitch and urea. 
For the force field of PreQ$_{1}$, we followed the basic philosophy of CHARMM general force field parameterization strategy \cite{Foloppe2000JCC}. For consistency with CHARMM force fields of canonical nucleotides, we applied partial charge from guanine to pyrimidine ring of PreQ$_{1}$ except for -C7-CH$_{2}$-NH$_{2}$ side chain and two carbon atoms directly connected to C7 atom. For -C7-CH$_{2}$-NH$_{2}$ and two connected carbon atoms, we assigned  partial charges and dihedral force constants by analogy using parameters from residues of similar structure in CHARMM force field as a first guess and refining the parameters in order to produce similar dihedral angle distribution to the preQ$_1$ in solution structure. 
For production runs, we performed simulations in various conditions, ranging from pure water to aqueous urea solution of differing concentrations. We simulated one 0 M-urea and three 8 M-urea each for 700 ns, two 4 M-urea each for 600 ns %three 2M-TMAO each for 700 ns 
in the presence of preQ$_{1}$ and Ca$^{2+}$, and one 0 M-urea for 700 ns and one 8 M-urea for 600 ns, in the absence of preQ$_{1}$ and Ca$^{2+}$. 
The comprehensive simulations provide fundamental insights into the mechanisms of destabilization of folded RNA. 
%We also simulated three 8 M-urea for 700 ns and one 0 M-urea for 600 ns in the presence of preQ$_{1}$ and without Ca$^{2+}$, and two 0 M-urea for 700 ns in the absence of preQ$_{1}$ and in the presence of Ca$^{2+}$.  

\section{Acknowledgements}
We thank Korea Institute for Advanced Study and Korea Institute of Science and Technology Information Supercomputing Center for providing computing resources.  
DT acknowledges support from the National Science Foundation through grant CHE 09-14033. \\

\noindent{\bf Supporting Information Available:}
Dynamics of RS upon removal of preQ$_1$ and Ca$^{2+}$ ions; Permeation of water into tertiary motifs;  Water, urea dynamics in the process of base pair rearrangement; Figures S1-S12; SI movie 1-5. This material is available free of charge via the Internet at http://pubs.acs.org.
\clearpage

%\bibliographystyle{achemso}
%\bibliography{bib,mybib1}
\ifx\mcitethebibliography\mciteundefinedmacro
\PackageError{achemso.bst}{mciteplus.sty has not been loaded}
{This bibstyle requires the use of the mciteplus package.}\fi

\clearpage
%\begin{landscape}

%fig1
%\end{landscape}

%fig2

%fig3

%fig4

%fig5

%fig6

%\begin{figure}
%\centering{\includegraphics[width=4.5in]{TOC_riboswitch}}
%\caption{TOC graphic
%\label{TOC}}
%\end{figure}
\clearpage 

\section{Supporting Information} 

\noindent{\bf Dynamics of RS upon removal of preQ$_{1}$ and Ca$^{2+}$ ions.}
Given that the biological function of preQ$_{1}$-RS is the translational gene regulation, which is dictated by the metabolite-dependent conformational changes involving secondary structure rearrangement \cite{AlHashimi08COSB,rieder2010PNAS}, how the RS responds to chemical stress is of great interest. 
Removal of the metabolite preQ$_{1}$ and Ca$^{2+}$ ions, which strongly stabilize the RS structure \cite{Zhang11JACS}, from their binding sites triggers a significant secondary structure rearrangements in the RS structure (Fig.S3a).  
%The time scale of full conversion between the two alternative RS structures, involving translation regulation, is beyond the limit of current all atom MD simulation. 
We also found that without preQ$_{1}$ and Ca$^{2+}$ the RS structure is dramatically destabilized in 8 M urea solution (Fig.S3b). 
RMSD of the A-rich tail (U25-A36) shows large fluctuations during the first half of simulation (0-300 ns) in the absence of preQ$_1$ and Ca$^{2+}$ but later reaches a stationary value by forming new base pairs. 
The tertiary contacts between the A-rich tail (U26$\sim$C33) and bases in the 5'-end, composed of Hoogsteen pairs (depicted by blue lines), are less stable than the secondary base pairs in the main stem-loop (see change of RMSD in Fig.S3a). 
Our simulations show that the structure surrounding preQ$_1$ and Ca$^{2+}$ ions is particularly stable compared with other regions. 
Hence, the melting of whole RS structure hinges on the tertiary structure surrounding the preQ$_1$ ligand.\\

\noindent{\bf Permeation of water into the tertiary motifs.}
Disruption of Hoogsteen base pairs that constitute pseudoknot stacks between the A-rich tail and the 5'-side of the sequence go through a dynamic process of water and urea, which is similar to the disruption of G4-C23 pair discussed in Fig.6 (see Figure \ref{si-13}). 
$n(O_{\rm w})$ and $n(O_{\rm u})$ calculated in the FSS and SSS around the H6$_{A27}$ show that disruption of N3$_{A5}$-H6$_{A27}$ hydrogen bond accompanies decrease of urea and increase of water at the early stage of disruption, which is later followed by an increase of both urea and water (190 ns $\sim$). 
It is worth noting that when the N3$_{A5}$-H6$_{A27}$ distance is increased to $\sim$ 5\AA\ ($\sim$ 190 ns), urea density starts increasing in the FSS. 
This indicates that water permeation further triggers urea-induced denaturation of tertiary interaction as well by perturbing base-base stack and hydrogen bonds of base pairs.  
When the disruption is completed, water population ceases to increase and urea density continues to grow until the unbound state of A27 is fully solvated by urea (195-205 ns). 
Together with the snapshots from the simulations over time, the interbase distance distributions (Figure \ref{si-13}(c)) quantify how the dynamics between bases evolves. 
To summarize, water permeation to the dry space between bases initially perturbs base pair hydrogen bonding, increasing the accessibility of urea to the bases. It is reported that water-nucleobase ``stacking" interactions exist which can enhance base-stacking interactions \cite{sarkhel2003}. Therefore, water might play a role in stabilizing base stacking until it is replaced by more stable urea-base stacking interactions.\\

\noindent{\bf Water, urea dynamics in the process of base pair rearrangement.} 
We examined the change in the number density of water/urea oxygens around base hydrogen during the process of G6-C21$\rightarrow$G6-A28 (secondary$\rightarrow$tertiary) rearrangement, as in the analyses of secondary and tertiary structure melting. Conformational changes in this process are shown in Figure \ref{si-14}. 
Water and urea densities in the first solvation shell of H1$_{G6}$ shows very clear pattern that water increases rapidly during melting of the base pairs, then decreases slowly after melting (curve with red circle). Urea increases continuously until melting is completed (curve with blue circle). 
Together with melting processes of base pairs in secondary and tertiary motifs, this again confirms  the microscopic mechanism that water permeates first to break base pair hydrogen bond, and urea stabilizes unbounded conformation.  
\\

%\bibliographystyle{achemso}
%\bibliography{bib,mybib1}
\ifx\mcitethebibliography\mciteundefinedmacro
\PackageError{achemso.bst}{mciteplus.sty has not been loaded}
{This bibstyle requires the use of the mciteplus package.}\fi

\newpage
%S1
\renewcommand{\thefigure}{S1}
\begin{figure}
\centering
\includegraphics[scale=0.35]{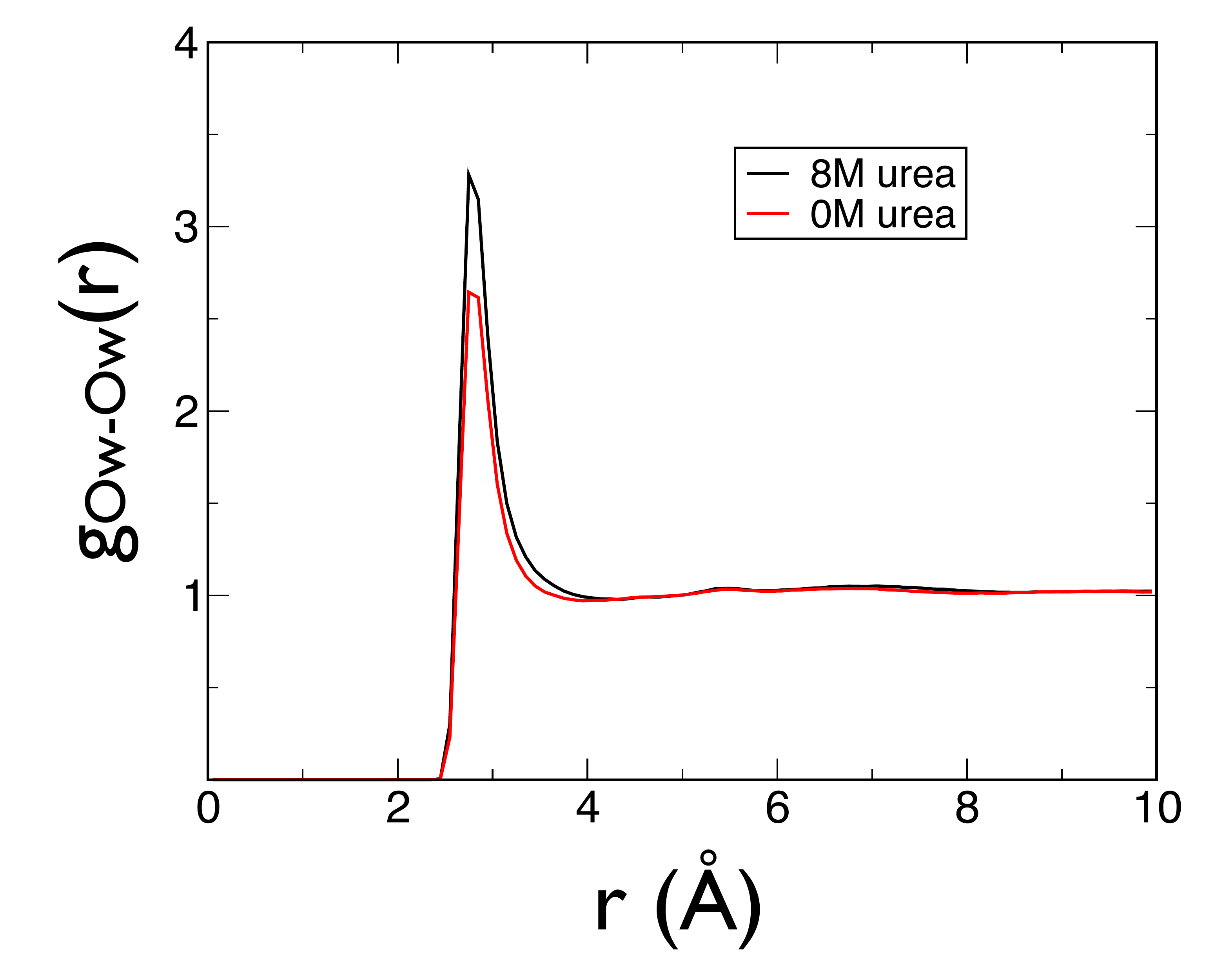} 
\caption{O-O radial distribution of water in 0M and 8M urea solution.
}
\label{si-1}
\end{figure}

%S2
\renewcommand{\thefigure}{S2}
\begin{figure}[h!] 
\centering
\includegraphics[scale=0.55]{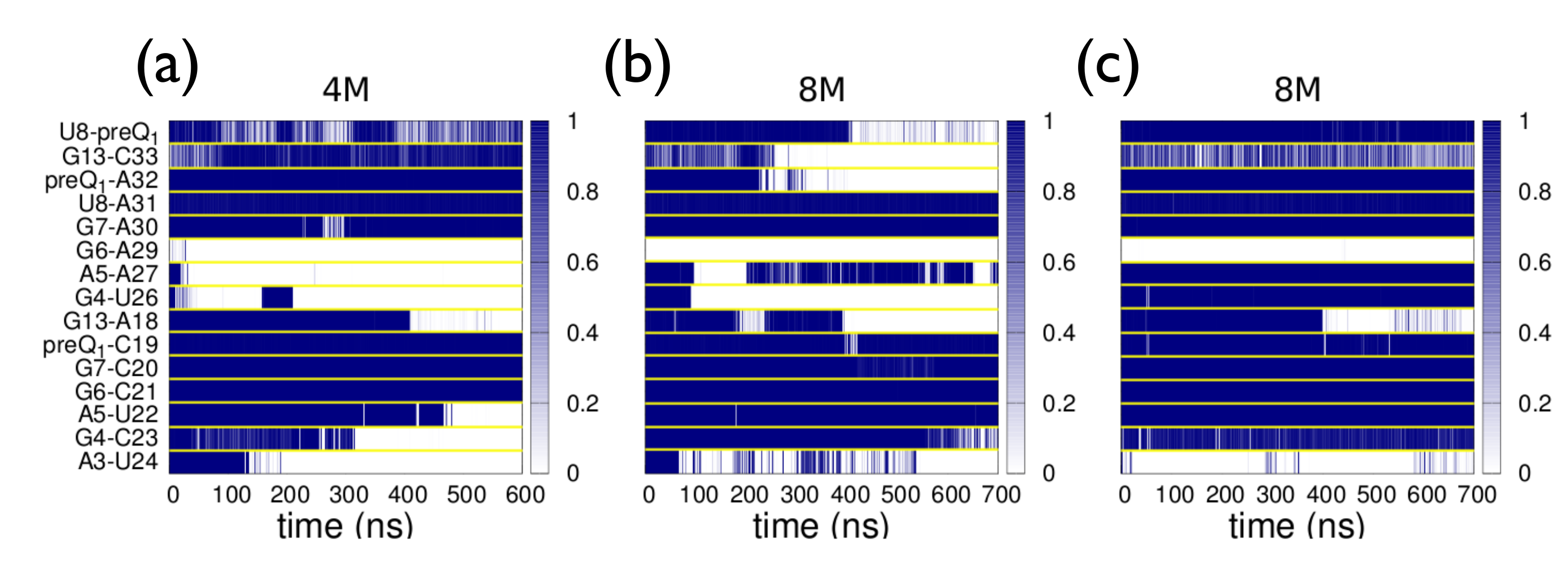}
\caption{Time evolution of base pair contacts for simulation trajectories in aqueous 4M and 8M urea solution. 
We use the logistic function $\varphi(r)$ defined in the main text (Eq.1).
}
\label{si-2}
\end{figure}

%S3
\renewcommand{\thefigure}{S3}
%\begin{landscape}
\begin{figure}
\includegraphics[scale=0.23]{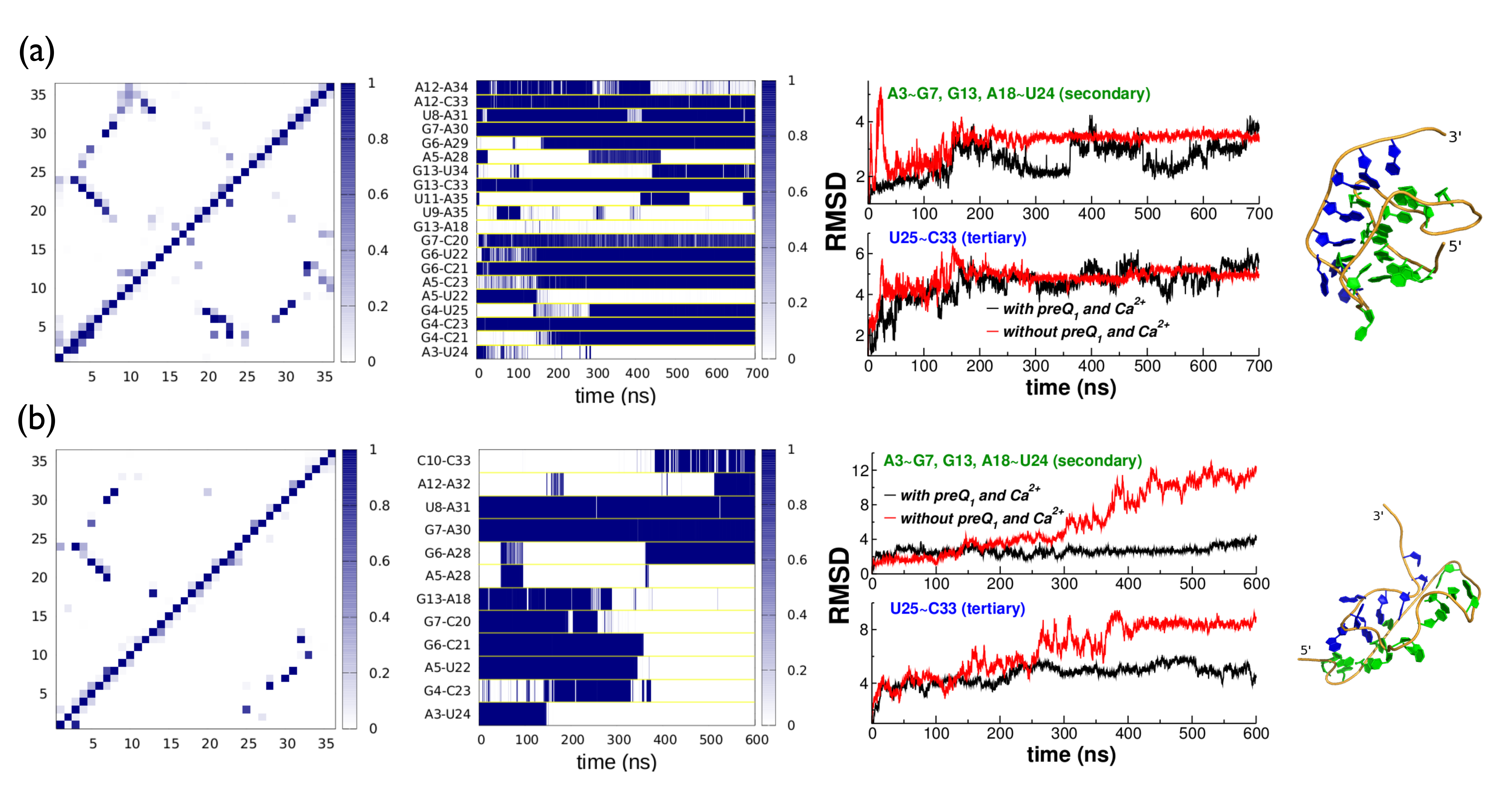}
\caption{Results from RS simulation without Ca$^{2+}$ and preQ$_{1}$ under {\bf (a)} urea-free and {\bf (b)} 8M urea condition. From left to right are shown base-base contact map at first 50 ns (upper left) and at last 50 ns (lower right), 
Time evolution of base pair contact, RMSDs of secondary and tertiary structure region of RS in comparison to a simulation trajectory in the presence of both Ca$^{2+}$ and preQ$_{1}$, and the conformation at the end of the simulation.
}
\label{preQ1_Ca_effect}
\end{figure}
%\end{landscape}

%S4
\renewcommand{\thefigure}{S4}
\begin{figure}[h] 
\begin{center}
\includegraphics[scale=0.45]{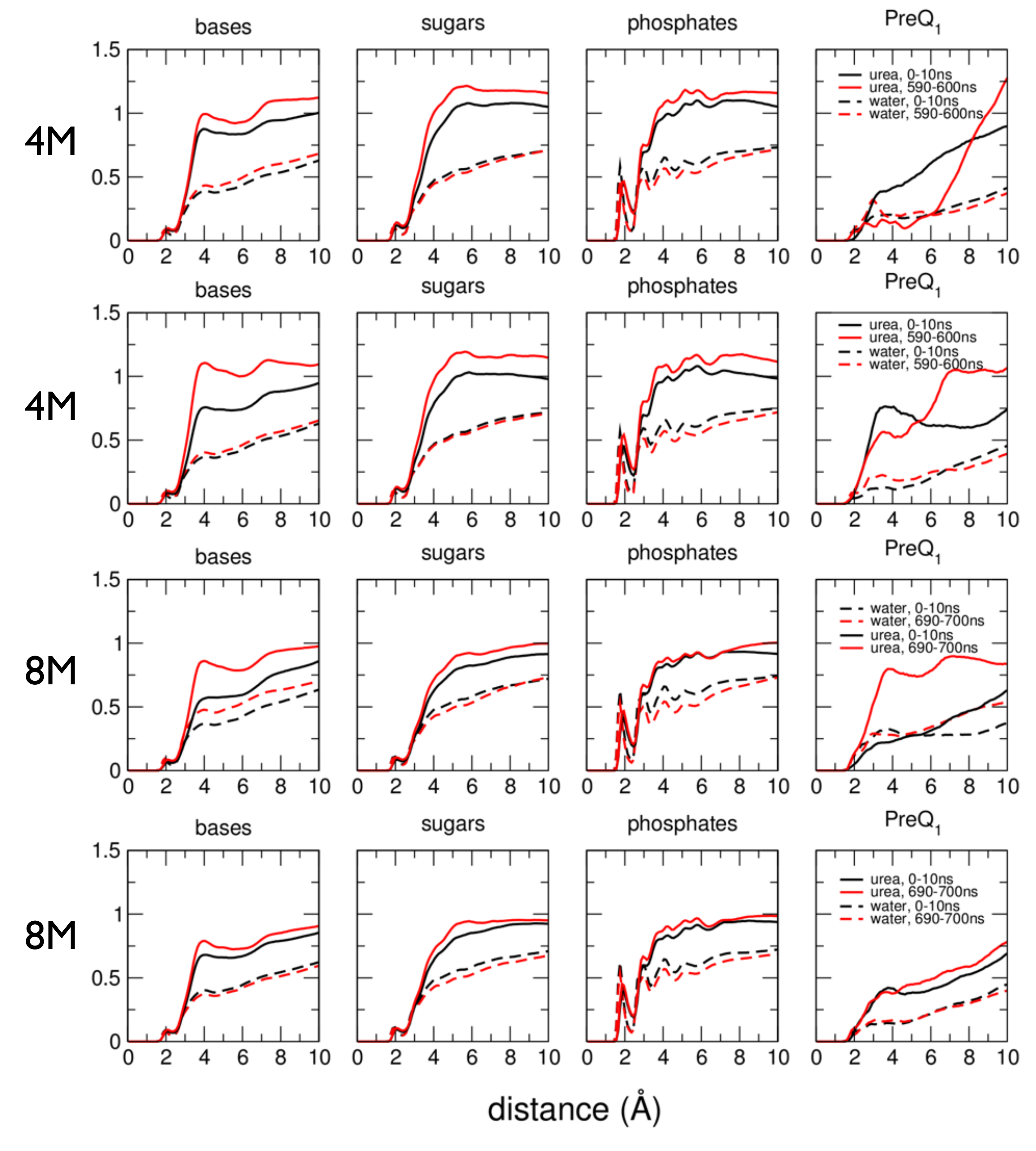}
\caption{Radial distribution functions (RDF) of water (dashed lines) and urea (solid lines) for bases, sugars, phosphates, and preQ$_1$ of RS in 4M (for the trajectories in Fig.1(c) and Fig.S2(a)) and 8M aqueous urea (for the trajectories in Figs.S2(b) and S2(c)). 
Here the distance for RDF is defined as the minimum distance from urea (or water) to any heavy atom in the specified group}
\end{center}
\label{si-4}
\end{figure}

%S5
\renewcommand{\thefigure}{S5}
\begin{figure}[ht]
\begin{center}
\includegraphics[scale=0.4]{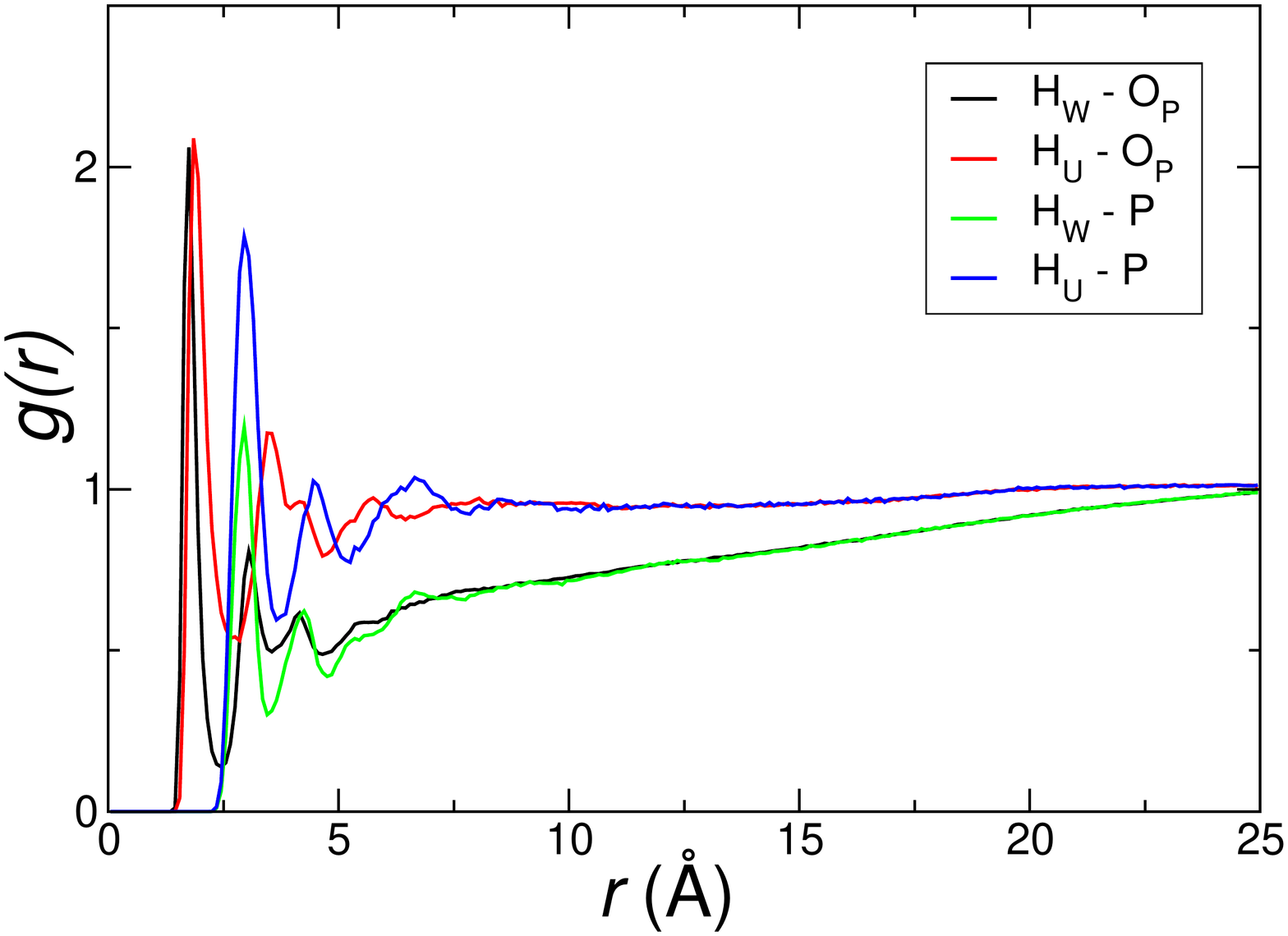} 
\end{center}
\caption{Radial distribution functions of water hydrogen (H$_{\rm W}$) - phosphate oxygen (O$_{\rm P}$), urea hydrogen (H$_{\rm U}$) - phosphate oxygen (O$_{\rm P}$), water hydrogen (H$_{\rm W}$) - phosphorus (P), urea hydrogen (H$_{\rm U}$) - phosphorus (P) in 8 M urea.We used last 10 ns of the trajectory. The graphs show clearly that the hydrogen atoms of water and urea make direct hydrogen bondings with phosphate oxygen.}
\label{si-5}
\end{figure}

%S6
%\renewcommand{\thefigure}{S6}
%\begin{figure} 
%\centering
%\subfloat[$\mathrm{g_u}(r|B_i)$]{\label{fig:a}%
%\includegraphics[scale=0.6,trim= 2cm 2cm 2cm 2cm,clip]{rdf-r-base-urea.pdf}}
%\hspace{.1cm}
%\subfloat[$\mathrm{g_w}(r|B_i)$]{\label{fig:b}%
%\includegraphics[scale=0.6,trim= 2cm 2cm 2cm 2cm,clip]{rdf-r-base-water.pdf}}
%\caption{Radial distribution functions of (a) urea and (b) water around each of 36 base in RS at first 10 ns (black line) and last 10 ns (red line) under 8M urea condition.}
%\label{si-6}
%\end{figure}

%S7
%\renewcommand{\thefigure}{S7}
%\begin{figure} 
%\centering
%\subfloat[$\mathrm{g_u}(r|S_i)$]{\label{fig:a}%
%\includegraphics[scale=0.6,trim= 2cm 2cm 2cm 2cm,clip]{rdf-r-sugar-urea.pdf}}
%\hspace{.1cm}
%\subfloat[$\mathrm{g_w}(r|S_i)$]{\label{fig:b}%
%\includegraphics[scale=0.6,trim= 2cm 2cm 2cm 2cm,clip]{rdf-r-sugar-water.pdf}}
%\caption{Radial distribution functions of (a) urea and (b) water around each of 36 ribose sugar in RS at first 10 ns (black line) and last 10 ns (red line) under 8M urea condition.}
%\label{si-7}
%\end{figure}

%S8
%\renewcommand{\thefigure}{S8}
%\begin{figure} 
%\centering
%\subfloat[$\mathrm{g_u}(r|P_i)$]{\label{fig:b}%
%\includegraphics[scale=0.6,trim= 2cm 2cm 2cm 2cm,clip]{rdf-r-bb-urea.pdf}}
%\hspace{.1cm}
%\subfloat[$\mathrm{g_w}(r|P_i)$]{\label{fig:a}%
%\includegraphics[scale=0.6,trim= 2cm 2cm 2cm 2cm,clip]{rdf-r-bb-water.pdf}}
%\caption{Radial distribution functions of (a) urea and (b) water around each of 36 phosphate backbone in RS at first 10 ns (black line) and last 10 ns (red line) under 8M urea condition.}
%\label{si-8}
%\end{figure}

\renewcommand{\thefigure}{S6}
\begin{figure} 
\includegraphics[scale=0.35]{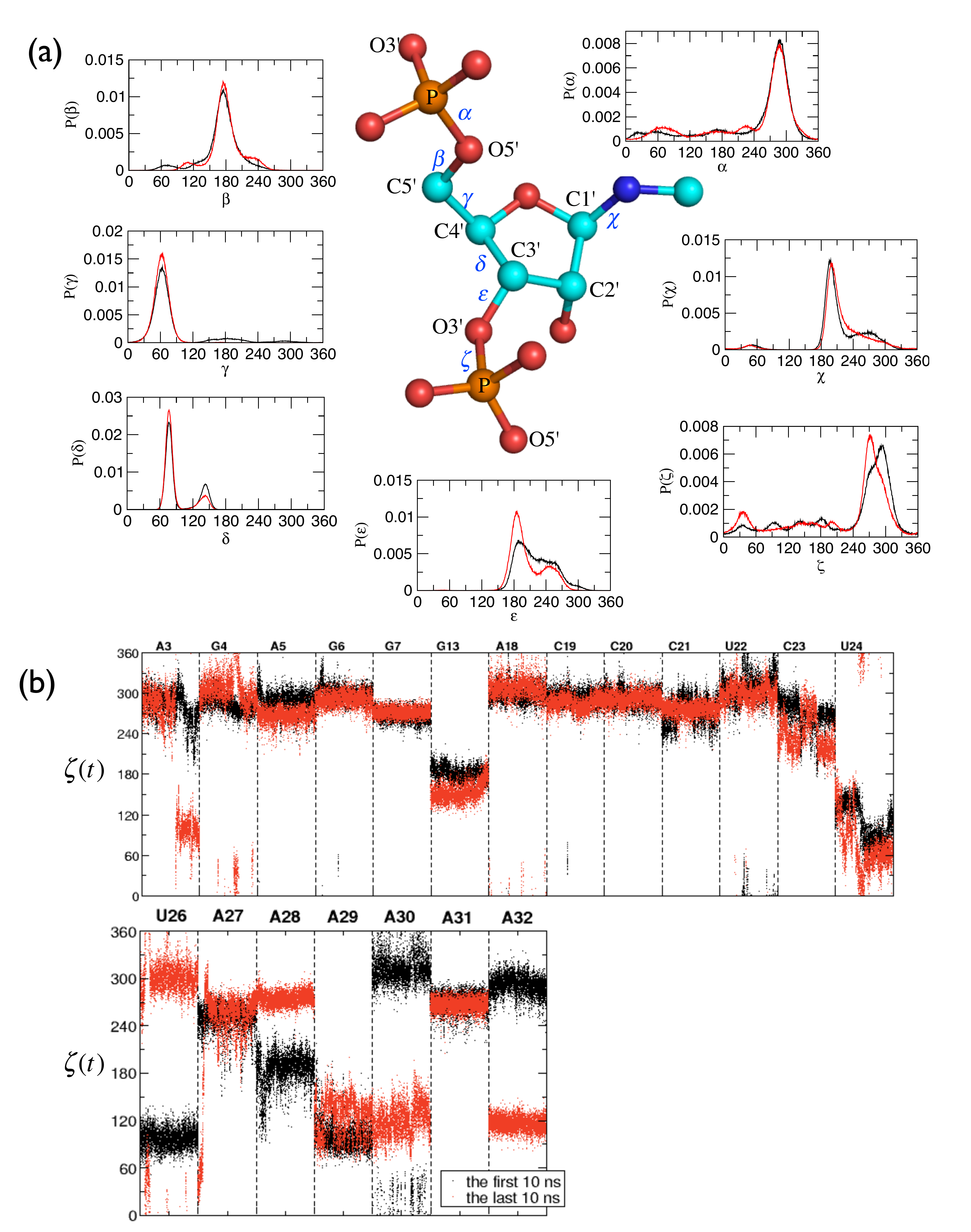}
\caption{Distributions of seven torsion angles ($x=\alpha$, $\beta$, $\gamma$, $\delta$, $\epsilon$, $\zeta$, $\chi$) along the phosphodiester backbone of nucleic acids before and after the urea-induced denaturation. 
(a) $P(x)$s from the first (black) and last (red) 10 ns simulation show how the angle distributions change. 
(b) Time trace of $\zeta$ angles of each nucleotide. The top and bottom panels are for the nucleotides in secondary and tertiary motifs, respectively. }
\label{si-9}
\end{figure}
\clearpage 

\renewcommand{\thefigure}{S7}
\begin{figure} 
\includegraphics[scale=0.35]{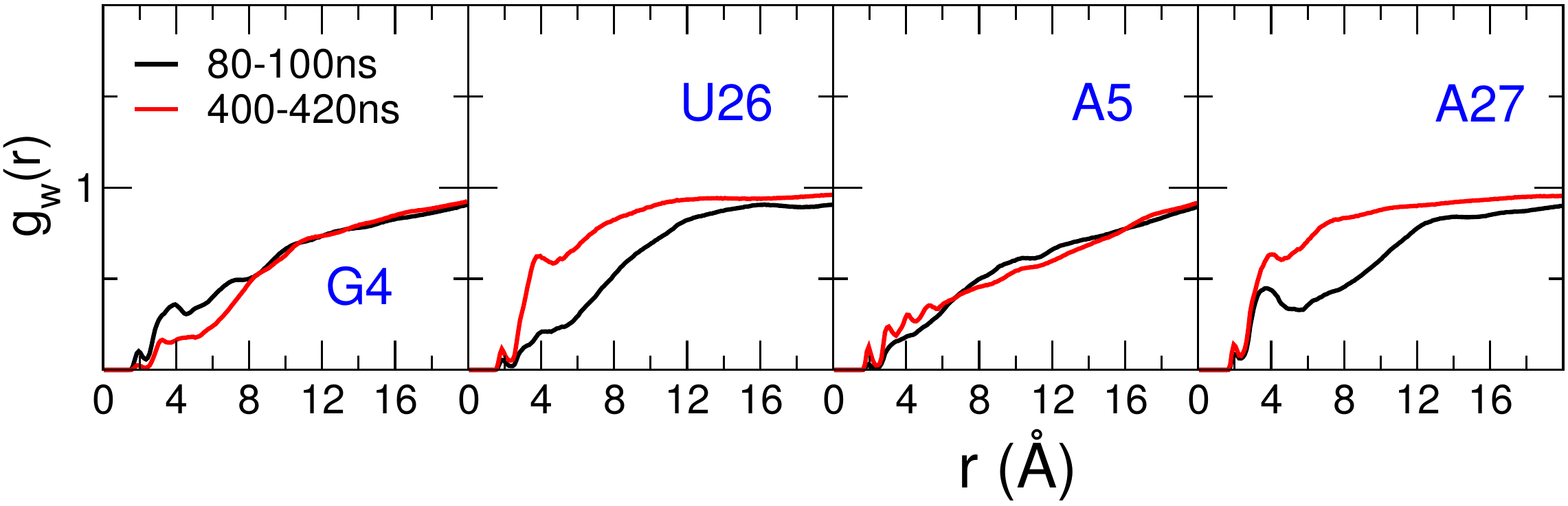}
\includegraphics[scale=0.35]{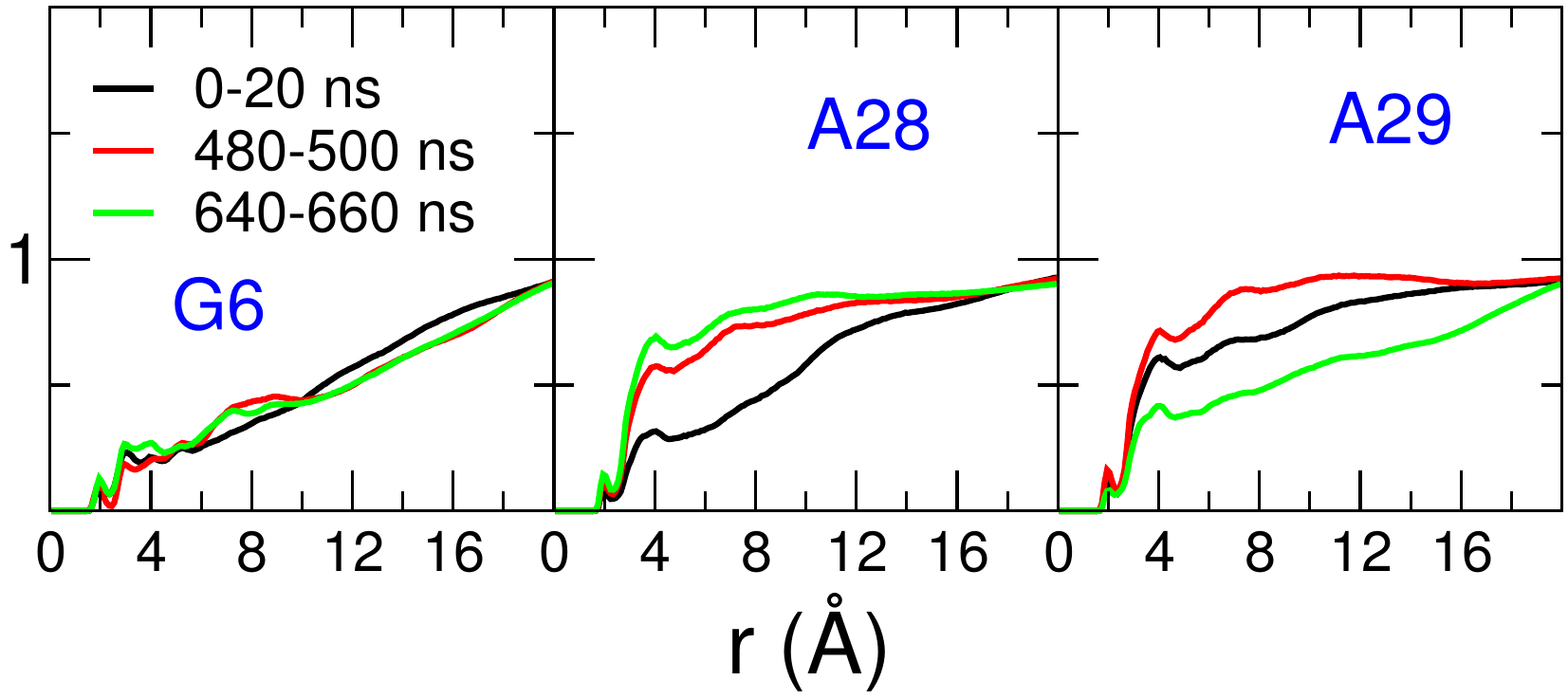}
\caption{$\mathrm{g_w}(r)$ around each base at the time intervals in Fig.3c}
\label{si-10}
\end{figure}

\renewcommand{\thefigure}{S8}
\begin{figure} 
\centerline{%
\includegraphics[scale=0.23]{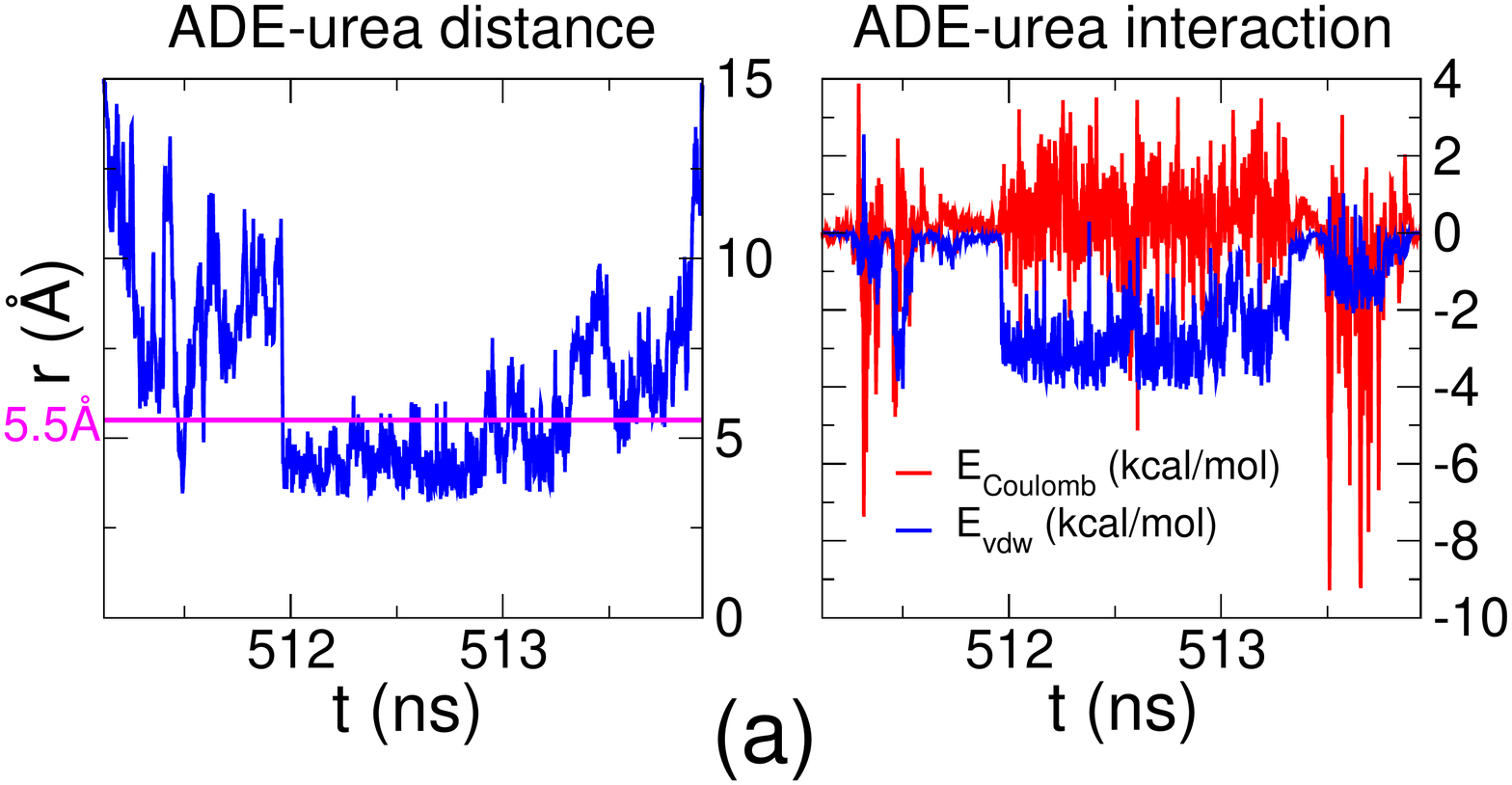}%
\includegraphics[scale=0.23]{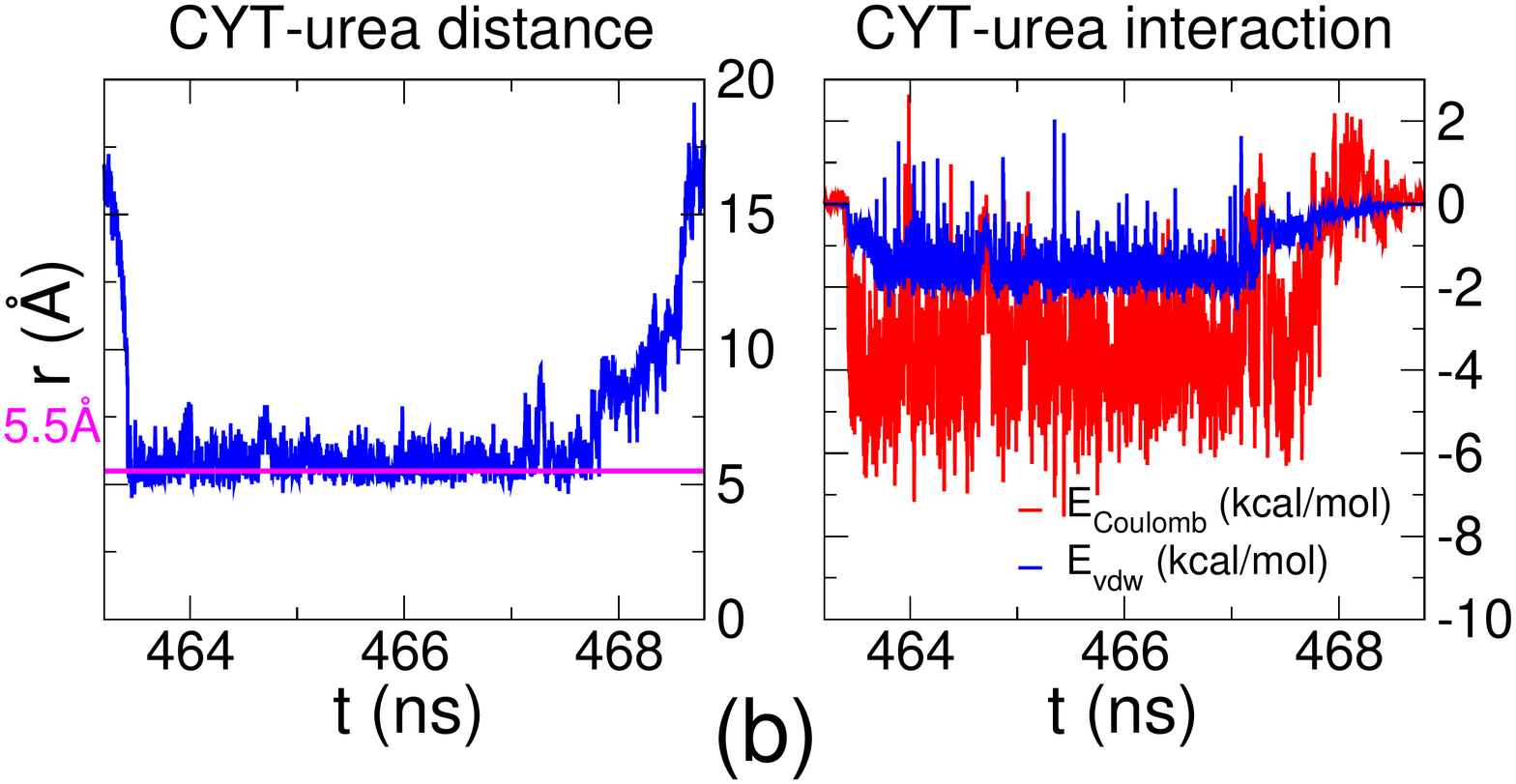}%
\includegraphics[scale=0.23]{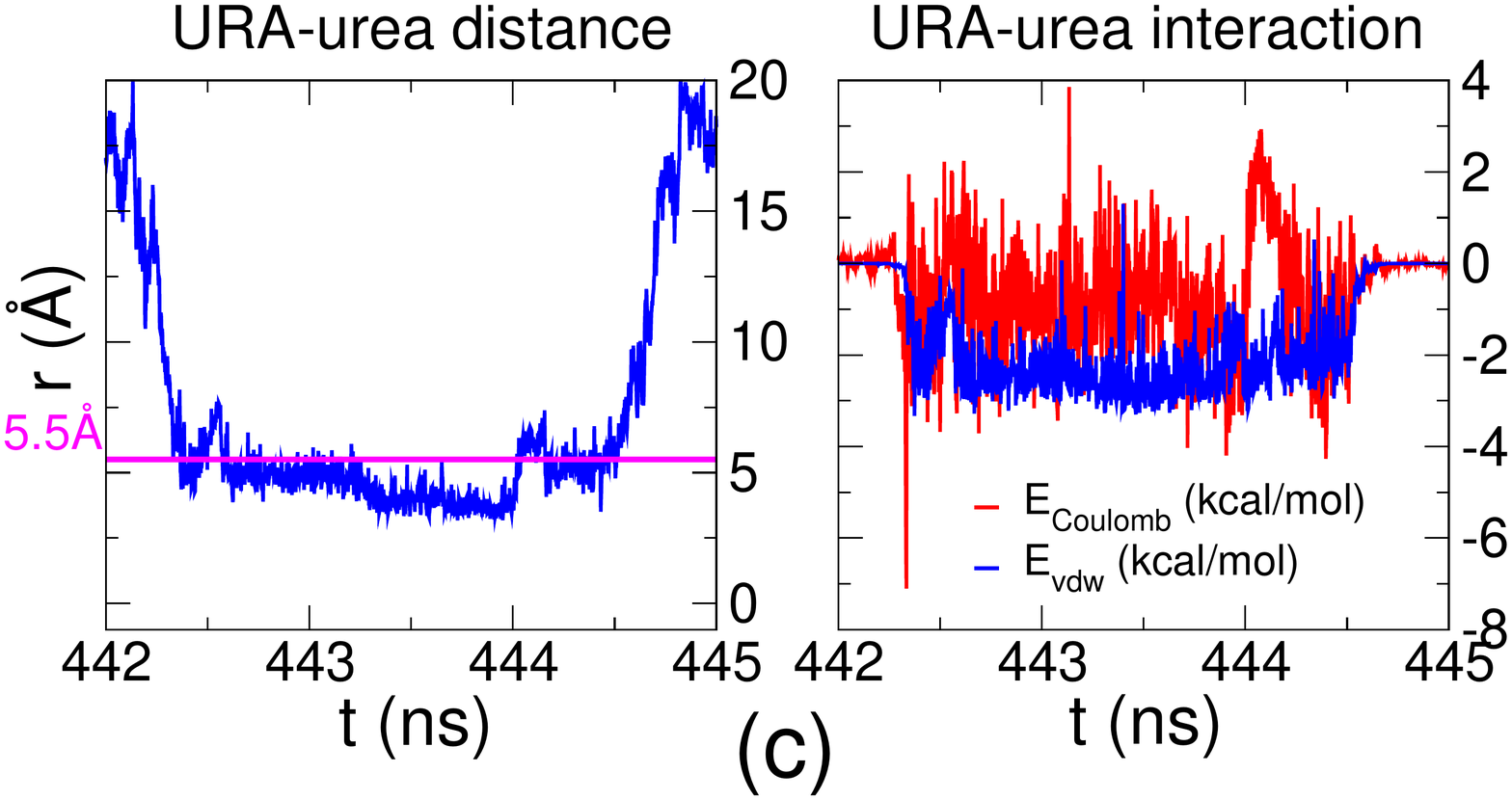}%
}%
\caption{vdW and electrostatic (Coulombic) energies of urea stack with adenine, cytosine, and uracile, respectively.}
\label{si-11}
\end{figure}

\renewcommand{\thefigure}{S9}
\begin{figure} 
\centerline{%
\includegraphics[scale=0.46]{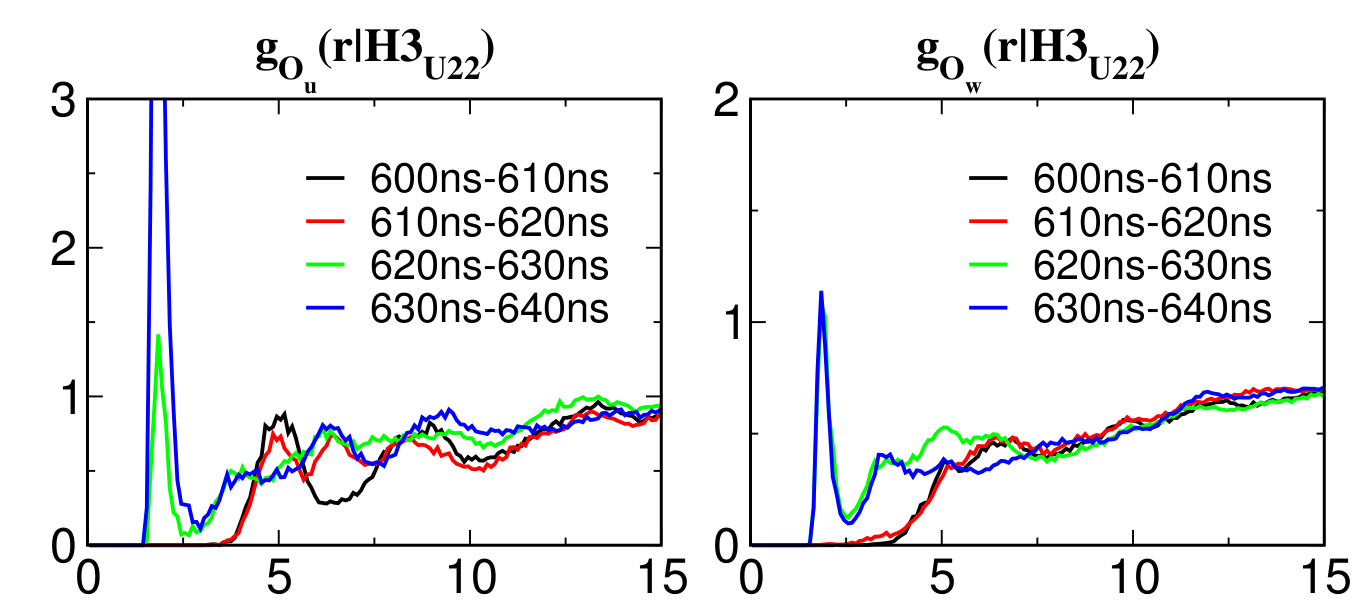}%
}%
\caption{Change in RDFs of urea oxygen and water oxygen around hydrogen atoms participating A5-U22 base pair interactions.}
\label{si-12}
\end{figure}

\clearpage

\renewcommand{\thefigure}{S10}
%\begin{landscape}
\begin{figure}
\centerline{%
\includegraphics[scale=0.40]{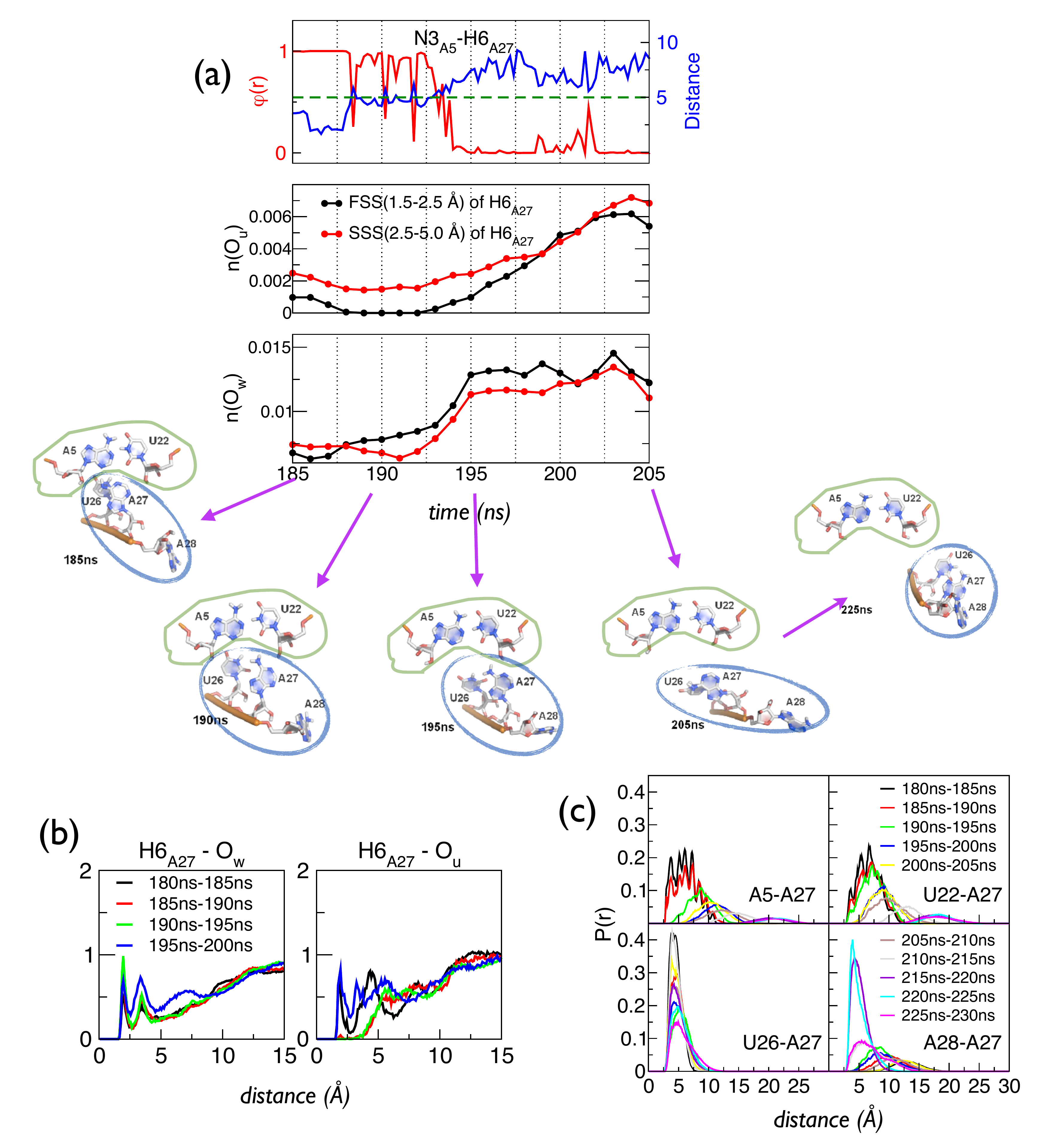}
}
%\includegraphics[scale=0.75,trim= 0cm 0cm 0cm 0cm,clip]{FigS12}%
%\includegraphics[scale=0.45,trim= 0cm 2cm 1cm 1cm,clip]{fig8b.png}%
%}%
%\centerline{%
%\textbf{(a)}\hspace{6cm}\textbf{(b)}%
%\hspace{8cm}\textbf{(c)}\hspace{0cm}%
%}%
\caption{Detailed dynamics of urea and water in the melting process of a Hoogsteen pair \textbf{(a)} (top) $\varphi(r)$ and distance $r$ between N3$_{A5}$ and H6$_{A27}$ atoms. 
(middle, bottom) Change of number densities of urea and water oxygens in the FSS and SSS of  H6$_{\rm A27}$ atom are calculated based on the time dependent RDF shown in (b). Series of snapshots over time show how the Hoogsteen pair between A5-A27 is being disrupted while the A5-U22 pair in the P1-stack stays intact. {\bf (b)} Change of RDF of O$_{\rm u}$ and O$_{\rm w}$ around H6$_{\rm A27}$ atom in the disruption dynamics of A5-A27 tertiary interaction. 
{\bf (c)} Pair distance distributions of A5-A27, U22-A27, U26-A27, and A28-A27 show how the interbase distance changes over time. }
\label{si-13}
\end{figure}
%\end{landscape}

\clearpage
\renewcommand{\thefigure}{S11}
\begin{figure}
\centerline{%
\begin{tabular}{cc}
\begin{tabular}{c}
\end{tabular} &
\begin{tabular}{c}
\includegraphics[scale=0.32,trim= 1cm 0.5cm 1cm 0.5cm,clip]{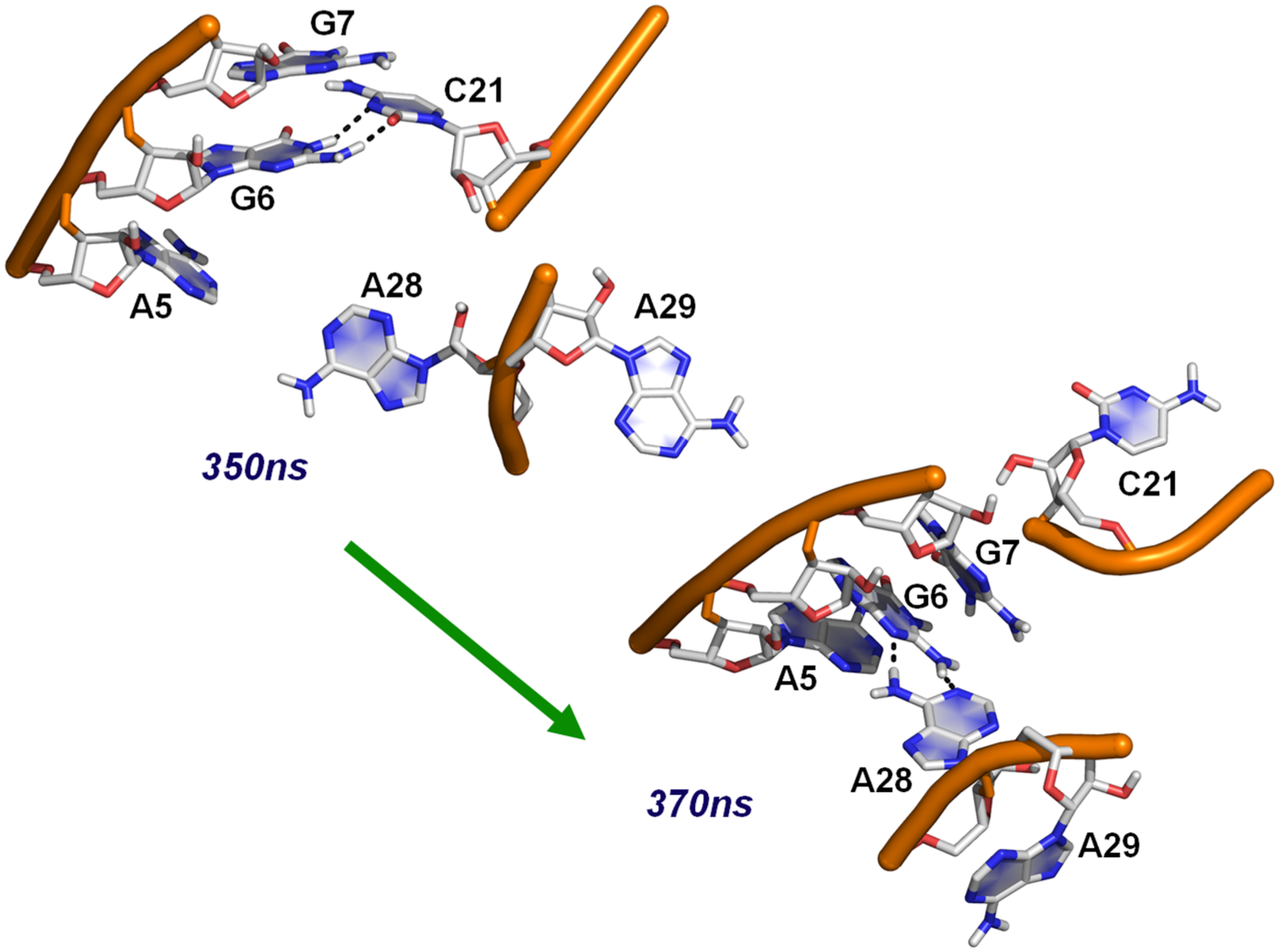}\\%
\includegraphics[scale=0.55,trim= 0.5cm 0cm 12cm 0cm]{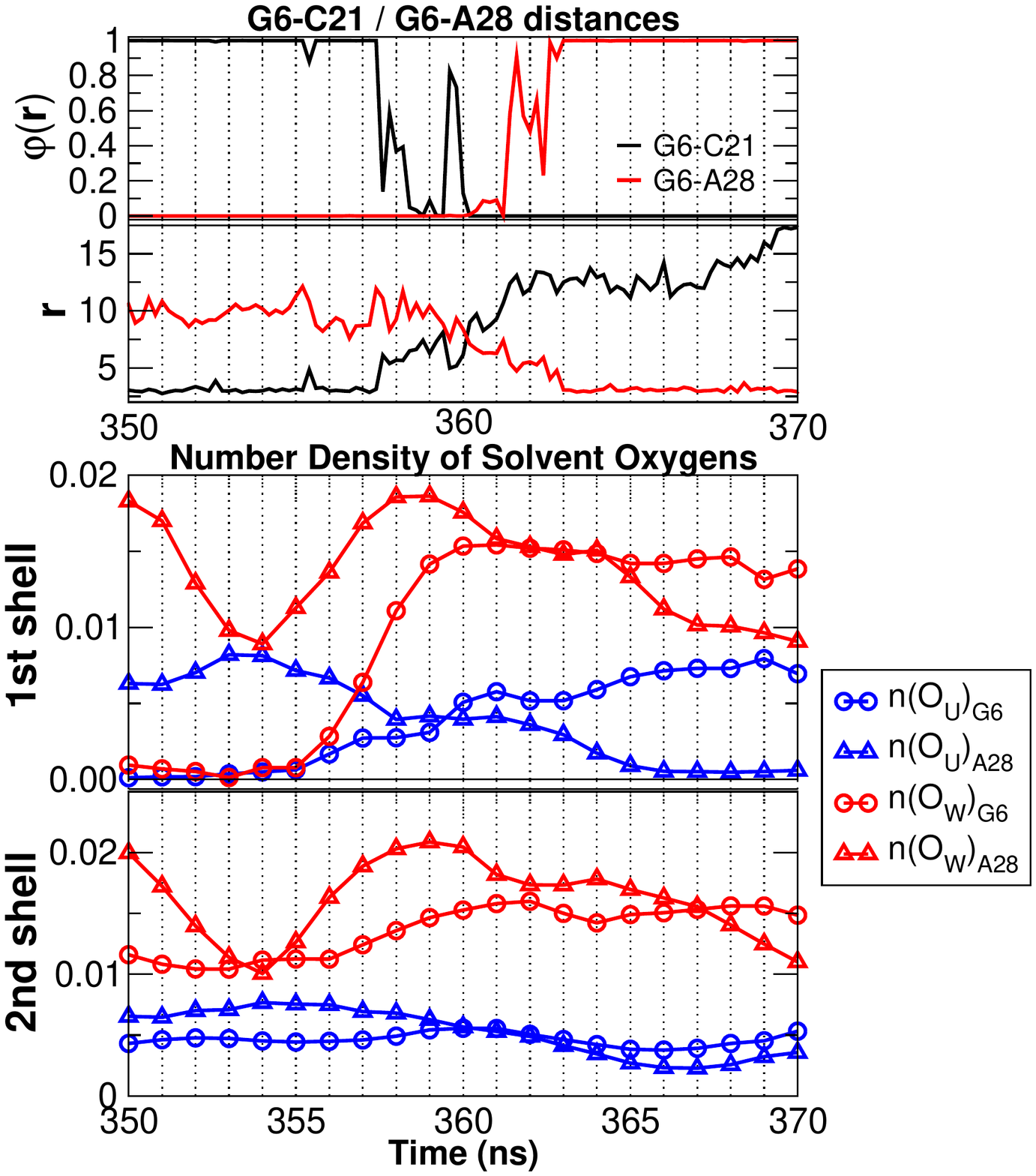}%
\end{tabular}
\end{tabular}
}%
%\centerline{\textbf{(a)}\hspace{8cm}\textbf{(b)}}%
\caption{Base pair rearrangement from G6-C21 to G6-A28. $\varphi(r)$ (Eq.1 in the main text) is computed between the hydrogen atoms of acceptor and donor bases.}
\label{si-14}
\end{figure}

\clearpage
\renewcommand{\thefigure}{S12}
\begin{figure}
\centering
\includegraphics[scale=0.52]{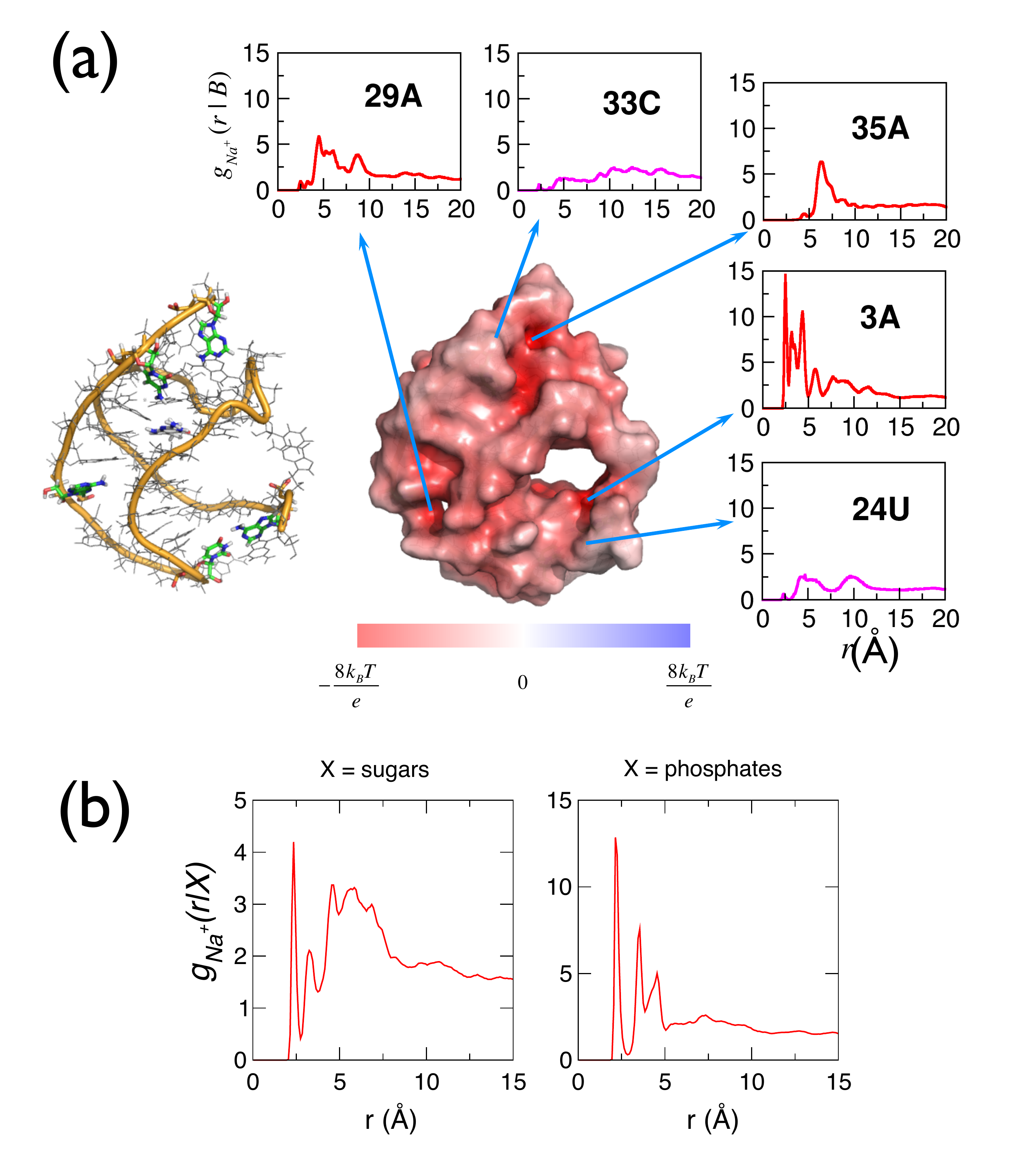}\\%
%\centerline{\textbf{(a)}\hspace{8cm}\textbf{(b)}}%
\caption{Distribution of counterion (Na$^+$) around the riboswitch structure.
(a) Electrostatic potential map of preQ$_1$-RS structure calculated at the solvent accessible surface by solving the nonlinear Poisson-Boltzmann equation. 
In accord with the map, radial distribution functions of Na$^+$ around several selected bases show that a larger amount of counterions (Na$^+$) are present near the more negatively charged surfaces.     
(b) All the radial distribution functions of Na$^+$ ions around sugar and phosphate group look similar regardless of the position of nucleotide in the RS structure. A high density of Na$^+$ ions in the vicinity of phosphate groups is noteworthy.}  
\label{counterion}
\end{figure}

\clearpage

%\renewcommand{\thefigure}{S14}
%\begin{figure}
%\centerline{%
%\includegraphics[scale=0.7,trim= 0cm 7cm 16cm 0cm,clip]{zeta.pdf}%
%\includegraphics[scale=0.7,trim= 0cm 7cm 16cm 0cm,clip]{zeta-tert.pdf}%
%}%
%\centerline{\textbf{(a)}\hspace{8cm}\textbf{(b)}}%
%\caption{Probability distributions of O5$^{\prime}$-P-O3$^{\prime}$-C3$^{\prime}$ dihedral angle($\zeta$) of phosphate backbone for first 10ns and last 10ns on each of three different simulation trajectories on preQ$_{1}$-RS under 8M urea condition \textbf{(a)} for all 36-nt of RS \textbf{(b)} for U26$\sim$C33 which forms tertiary interactions.}
%\label{si-13}
%\end{figure}

%\clearpage
%\renewcommand{\thefigure}{S15}
%\begin{figure}
%\centerline{%
%\includegraphics[scale=0.7,trim= 0.5cm 7cm 19.5cm 0cm,clip]{epsilon.pdf}%
%\includegraphics[scale=0.7,trim= 0.5cm 7cm 19.5cm 0cm,clip]{alpha.pdf}%
%\includegraphics[scale=0.7,trim= 0.5cm 7cm 19.5cm 0cm,clip]{chi.pdf}%
%}%
%\centerline{\textbf{(a)}\hspace{4cm}\textbf{(b)}\hspace{4cm}\textbf{(c)}}%
%\caption{Probability distributions of \textbf{(a)} dihedral angle $\varepsilon$(P-O3$^{\prime}$-C3$^{\prime}$-C4$^{\prime}$), \textbf{(b)} dihedral angle $\alpha$(C5$^{\prime}$-O5$^{\prime}$-P-O3$^{\prime}$), and \textbf{(c)} dihedral angle $\chi$(O4$^{\prime}$-C1$^{\prime}$-N1-C2 for C and U, and O4$^{\prime}$-C1$^{\prime}$-N9-C4 for A and G) of phosphate backbone for first 10ns and last 10ns on each of three different simulation trajectories on preQ$_{1}$-RS under 8M urea condition.}
%\label{si-15}
%\end{figure}

\noindent {\bf Captions for SI movie 1-5}\\

\noindent {\bf SI movie 1:} Dynamics of stacking interaction between urea and guanine base. This movie visualizes the simulation trajectory in Figure 4c, showing how a urea molecule approaches the base, maintains stacking interaction for about 8 ns, and subsequently diffuses away. 
\href{http://newton.kias.re.kr/~hyeoncb/homepage/publication/Urea_movie1.mpg}{{\color{blue}Download - SI movie 1}}
\\ 

\noindent {\bf SI movie 2:} Dynamics of urea on cytosine base. Because the area of base ring is small, urea-base stacking interaction is not as effective as with purine bases. Urea spends most of the time interacting with phosphate group or ribose, which makes urea-cytosine base stacking incomplete.
\href{http://newton.kias.re.kr/~hyeoncb/homepage/publication/Urea_movie2.mpg}{{\color{blue}Download - SI movie 2}}
 \\

\noindent {\bf SI movie 3:} Dynamics of urea on uracil base. Because the area of base ring is small, urea-base stacking interaction is not as effective as with purine bases. 
\href{http://newton.kias.re.kr/~hyeoncb/homepage/publication/Urea_movie3.mpg}{{\color{blue}Download - SI movie 3}}
\\

\noindent {\bf SI movie 4:} Dynamics of hydrogen bond interaction between urea and guanine base. This movie visualizes the simulation trajectory in Figure 5b. 
\href{http://newton.kias.re.kr/~hyeoncb/homepage/publication/Urea_movie4.mpg}{{\color{blue}Download - SI movie 4}}
\\

\noindent {\bf SI movie 5:} Dynamics of water penetration into the narrow space between base pair. This movie visualizes how the  water molecules represented with water oxygens increase in the FSS ($<3$ \AA, yellow spheres) and SSS ($<5$ \AA, green spheres) during the time interval of 596-601 ns in Figure 6c. Urea oxygens within 7 \AA\ from H1$_{G4}$ (magenta sphere) are depicted with red spheres.   
Urea oxygen form hydrogen bond with H1$_{G4}$ only after the water molecules fills up the space between G4-C23 base pairs. 
\href{http://newton.kias.re.kr/~hyeoncb/homepage/publication/Urea_movie5.mpg}{{\color{blue}Download - SI movie 5}}

\end{document}